\documentclass[%
preprint,
 amsmath,amssymb,
 aps,
pra,
floatfix,
]{revtex4-1}

\usepackage{graphicx}
\usepackage{dcolumn}
\usepackage{bm}
\usepackage{color}

\begin{document}


\title{Rovibronic spectra of molecules dressed by light fields}

\author{Tam\'as Szidarovszky}
\email{tamas821@caesar.elte.hu}
\affiliation{%
Laboratory of Molecular Structure and Dynamics, Institute of Chemistry,
ELTE E\"otv\"os Lor\'and University and MTA-ELTE Complex Chemical Systems Research Group,
H-1117 Budapest, P\'azm\'any P\'eter s\'et\'any 1/A, Hungary}%

\author{G\'abor J. Hal\'asz}
\affiliation{Department of Information Technology, University of Debrecen,
PO Box 400, H-4002 Debrecen, Hungary}

\author{Attila G. Cs\'asz\'ar}
\affiliation{Laboratory of Molecular Structure and Dynamics, Institute of Chemistry,
ELTE E\"otv\"os Lor\'and University and MTA-ELTE Complex Chemical Systems Research Group,
H-1117 Budapest, P\'azm\'any P\'eter s\'et\'any 1/A, Hungary}
	
\author{\'Agnes Vib\'ok}
\email{vibok@phys.unideb.hu}
\affiliation{Department of Theoretical Physics, University of Debrecen, PO Box 400,
H-4002 Debrecen, Hungary and ELI-ALPS, ELI-HU Non-Profit Ltd., Dugonics t\'er 13,
H-6720 Szeged, Hungary}	

\date{\today}

\begin{abstract}
The theory of rovibronic spectroscopy of light-dressed molecules is presented within
the framework of quantum mechanically treated molecules interacting with classical light fields.
Numerical applications are demonstrated for the homonuclear diatomic molecule Na$_2$,
for which the general formulae can be simplified considerably and the physical processes
leading to the light-dressed spectra can be understood straightforwardly.
The physical origin of different peaks in the light-dressed spectrum of Na$_2$ is given
and the light-dressed spectrum is investigated in terms of its dependence on the
dressing field's intensity and wavelength, the turn-on time of the dressing field,
and the temperature.
The important implications of light-dressed spectroscopy on deriving
field-free spectroscopic quantities are also discussed. 
\end{abstract}

\pacs{Valid PACS appear here}
\maketitle


\section{\label{Introduction}Introduction}

Atomic and molecular spectroscopy are among the most successful tools of science
both in fundamental research and in practical applications.
Despite the 200-year-long history of spectroscopy,
new approaches and methods are being developed to this day \cite{11MeQu}.
This is strongly related to the remarkable advances in the available
experimental techniques and light sources.
Some notable developments include frequency-comb
techniques \cite{Udem2002_freq_combs_1,Diddams2010_freq_combs_2,06Hall,06Hansch},
facilitating extremely precise and accurate measurements in the frequency domain,
and the ability to generate ultrashort and intense laser
pulses \cite{Krausz_fs_lasers,Krausz_as_lasers},
allowing for time-resolved spectroscopies on femtosecond or even attosecond \cite{Krausz_as_lasers,19PaMa} timescales.

A common approach in the application of spectroscopic techniques is the
use of two (or more) light pulses, with some pulses acting as so-called pump pulses,
which induce specific changes in the system, while subsequent, so-called probe pulses
are used (directly or indirectly) to measure the changes induced by the pump pulse(s).
If the duration of both the pump and the probe pulses is short, then repeating the experiment
with varying time delays between the pulses can lead to time-resolved dynamical
information \cite{Zewail2000_pumpprobe,Dantus2001_pumpprobe} or by means of a
Fourier transformation
to multidimensional and/or
high-resolution spectra \cite{Mukamel2000_multidim_spectroscopy,FT_cation_spectroscopy_PRL}.
If the pump and probe pulses overlap in time, the signal recorded by the probe pulse represents the so-called field-dressed
or light-dressed properties of the system \cite{CohenTannoudji,Shirley_Floquet_2,George_AccChemRes_1977},
where the pump pulse acts as the dressing field.
If both the pump and probe pulses are long with respect to the timescales
of the processes investigated,
then one obtains static spectral properties of the light-dressed system.
We investigate the last scenario in this work and henceforth call it
\emph{light-dressed spectroscopy}.
The theoretical and experimental methods for investigating spectral
transitions between atomic light-dressed states is well developed \cite{CohenTannoudji}. As to molecules, the concept of light-dressed electronic states, also called light-dressed potentials, has been utilized with success to understand nuclear dynamics both in experimental and theoretical studies \cite{Wunderlich1,LIP_Garraway_PRL_1998,LIP_Kono_JACS_2003,LIP_Corrales_NatChem_2014,Atabek_PRL_2009,Atabek2010,Bucksbaum3_PRL_1990}.
To some extent light-dressed rovibrational spectroscopy has been adopted for molecular systems, as well; for example, inducing Autler--Townes-type splittings \cite{AutlerTownes_original} of rotational transitions with microwave radiation have been used to deduce molecular parameters \cite{AutlerTownes_in_spectroscopy_2017,AutlerTownes_in_spectroscopy_2012} or to promote the assignment of rovibrational spectra \cite{AutlerTownes_in_spectroscopy_1997}. Related to light-dressed spectroscopy, the optical absorption of electronic materials driven away from equilibrium by laser light was also investigated \cite{spectra_of_laser_driven_materials_PRA_2018}.

In a previous theoretical study of the present authors, in which all molecular degrees of freedom were incorporated into the concept of light-dressed states, 
the rovibronic spectrum of light-dressed Na$_2$ was
investigated in the context of how the presence of a
Light-Induced Conical Intersection (LICI) \cite{LICI1,LICI2}, generated by the
dressing field, can be identified in the
spectrum \cite{Szidarovszky2018_JPCL_1,Szidarovszky2018_JPCL_2}.
The modeling work was carried out for both the classical dressing field of laser
radiation \cite{Szidarovszky2018_JPCL_1} and the quantized dressing field of
a microscopic cavity mode \cite{Szidarovszky2018_JPCL_2}.

It is our firm belief that the spectroscopy on light-dressed molecules is of fundamental interest and could provide valuable scientific results on both light-dressed and field-free molecular systems.
For example, because the response of light-dressed systems to external fields is different from that of the corresponding field-free systems, understanding and predicting the absorption and emission properties of light-dressed molecules can be useful for tuning or controlling optical processes, such as those in coherent control schemes or laser cooling \cite{coherent_control_Warren_Science_1993,coherent_control_Koch_ChemRev_2012,STIRAP,Atabek_vib_cooling_PRL_2011,optical_cooling_Viteau2008}.
In this work we provide an
introduction to the theory of computing
light-dressed spectra induced by classical light fields and investigate certain
aspects and spectroscopic implications of light-dressed spectroscopy, such as the effects of dressing-light intensity and frequency and those of finite non-zero temperature and the turn-on time of the dressing field. A procedure how to derive field-free transition frequencies from light-dressed spectra is also proposed.

\section{\label{Theory}The theoretical approach}
To make reading of this paper easier, the presentation of Floquet theory \cite{Floquet,Shirley_Floquet_2,Chu1985_Floquet_3} and
some of the detailed derivations have been moved to the Appendix, 
Section \ref{Appendix} (\ref{Appendix_A}, \ref{Appendix_B}, and \ref{Appendix_C})
of this paper.
The Appendix contains a large number of equations and only a few equations
are given in this section.
Those not familiar with Floquet 
theory or those
interested in modeling details may want to read the Appendix before reading
this section.

We advocate the use of a three-step theoretical approach to compute the light-dressed spectra of molecules. First, compute the field-free rovibronic states of the investigated molecule, second, using the field-free eigenstates as basis functions compute the light-dressed states, and finally, compute the transitions between the light-dressed states.  

As to the physical scenario to be simulated, we make a couple of important
assumptions:
(1) Initially the molecule is in a field-free eigenstate in the gas phase.
(2) The molecule is exposed to a medium intensity dressing light
($I = 10^7 - 5\cdot10^9$ Wcm$^{-2}$ in the numerical examples of this work),
which is turned on adiabatically, \emph{i.e.},
its envelope varies much slower than the rovibronic timescales characterizing the molecule
(such a dressing pulse converts the molecular wave function into a light-dressed
state or a superposition of light-dressed states, see Sections \ref{Appendix_A} and \ref{Appendix_B}).
In the numerical examples considered below, the slowest molecular timescale is around 100 ps; thus, dressing pulses of at least nanosecond length could be considered as turning on adiabatically.
(3) The probe pulse, introduced to record the static rovibronic spectrum
of the light-dressed molecule, is weak.

\subsection{Light-dressed states}
    
We determine the light-dressed states generated by the dressing light
within the framework of Floquet theory \cite{Floquet,Shirley_Floquet_2,Chu1985_Floquet_3}.
As detailed in Section \ref{Appendix_A},
in the presence of a dressing field periodic in time, the solution of the 
time-dependent Schr\"odinger equation (TDSE)
can be written as a superposition of light-dressed states (Floquet states)
$\vert \Phi _k (t) \rangle$.
With the initial wave function being a field-free eigenstate and the dressing-field
being turned on adiabatically, the generated light-dressed wave function is composed
of a single light-dressed state (see Sec. \ref{Appendix_B}) \cite{AdiabaticFloquet2_Gurin_AdvChemPhys_2003,AdiabaticFloquet_Lefebvre_PRA_2013}.
Because the $\vert \Phi _k (t) \rangle$ states are periodic in time,
they can be expanded as a Fourier series,
    \begin{equation}
        \vert \Phi _{k} \rangle = \sum_{n, \alpha,v,J} C^{(k)}_{n,\alpha v J} \vert \alpha v J\rangle \vert n \rangle,
        \label{eq:Floquet_state_expansion_general_text_body}
    \end{equation}
where we have used the notation called ``\textit{Floquet-state nomenclature}" \cite{Floquet},
in which $\langle t \vert n\rangle = e^{in \omega_1 t}$ and $\omega_1=2\pi/T$ 
with $\vert \Phi _k (t+T) \rangle=\vert \Phi _k (t) \rangle$.
In Eq. (\ref{eq:Floquet_state_expansion_general_text_body}), $\alpha$, $v$, and $J$
represent electronic, vibrational, and rotational quantum numbers, respectively.
The expansion coefficients $C^{(k)}_{n,\alpha v J}$ are obtained by diagonalizing
the Floquet Hamiltonian of Eq. (\ref{eq:HF_detailed}).

In practical applications, when the Floquet Hamiltonian can be simplified to the form of
Eq. (\ref{eq:HF_detailed_2x2}), \textit{i.e.},
when molecules have no permanent dipole and no intrinsic nonadiabatic couplings 
(or when the transitions originating from the permanent dipole and intrinsic nonadiabatic couplings can be neglected), 
and the dressing-field intensity is moderate enough to allow for the two-by-two Floquet Hamiltonian approach \cite{Halasz2012},
light-dressed states have a simplified form of 
    \begin{equation}
        \vert \Phi _{k} (n)\rangle = \sum_{v,J} C^{(k)}_{{\rm X} v J} \vert {\rm X} v J\rangle \vert n \rangle + \sum_{v,J} C^{(k)}_{{\rm A} v J} \vert {\rm A} v J\rangle \vert n-1 \rangle,
        \label{eq:Floquet_state_expansion_2x2_text_body}
    \end{equation}
where X and A represent the two electronic states considered (as usual, X denotes
the ground electronic state).

\subsection{Transitions between light-dressed states}
    
Once the light-dressed states have been determined,
one can compute the transition probabilities between the different
light-dressed states as induced by the weak probe pulse.
Following the standard approach of molecular spectroscopy \cite{98BuJe},
we use first-order time-dependent perturbation theory (TDPT1),
as detailed in Section \ref{Appendix_C}. 
    
For physical scenarios in which the light-dressed states have the form
shown in Eq. (\ref{eq:Floquet_state_expansion_2x2_text_body}),
such as the light-dressed states of the Na$_2$ molecule investigated below,
the $T_{l \leftarrow k}$ TDPT1 transition amplitude induced between
the \textit{k}th and \textit{l}th light-dressed states by an interaction of the form
$\hat{W}_2 (t) = - \mathbf{E}_2 \hat{\boldsymbol{\mu}} \, {\rm cos}(\omega _2 t)$ 
gives
(see Eq. (\ref{eq:transition_amplitude_Floquet_homonucleardiatomic}))
    \begin{equation}
        \begin{split}
            T_{l \leftarrow k} \propto \\
            \sum_{v,J} \sum_{v',J'} C^{(l)*}_{{\rm A} v' J'} C^{(k)}_{{\rm X} v J} \langle {\rm A} v' J' \vert \mathbf{E}_2 \hat{\boldsymbol{\mu}} \vert {\rm X} v J\rangle \delta(\varepsilon _k - \varepsilon _l - \hbar \omega_1 \pm \hbar \omega_2 ) + \\ 
            \sum_{v,J} \sum_{v',J'} C^{(l)*}_{{\rm X} v' J'} C^{(k)}_{{\rm A} v J} \langle {\rm X} v' J' \vert \mathbf{E}_2 \hat{\boldsymbol{\mu}} \vert {\rm A} v J\rangle \delta(\varepsilon _k - \varepsilon _l + \hbar \omega_1 \pm \hbar \omega_2 ),
        \end{split}
        \label{eq:Floquet_state_transition_amplitude_2x2_text_body}
    \end{equation}
where $\mathbf{E}_2$ is the electric field vector of the probe pulse,
$\mathbf{\mu}$ is the molecular dipole operator, $\hbar\omega_1$ and $\hbar\omega_2$
are the photon energies of the dressing and probe fields, respectively,
and $\varepsilon_i$ is the quasienergy of the \textit{i}th light-dressed state. 
    
Based on the arguments of the delta functions in 
Eq. (\ref{eq:Floquet_state_transition_amplitude_2x2_text_body}),
the first term can be interpreted as a transition between light-dressed states
having quasienergies $\varepsilon_k$ and $\varepsilon _l + \hbar \omega_1$, that is,
a transition between $\vert \Phi _{k} (n)\rangle$ and $\vert \Phi _{l} (n')\rangle$
with $n=n'-1$.
Similarly, the second term in 
Eq. (\ref{eq:Floquet_state_transition_amplitude_2x2_text_body})
can be interpreted as a transition between $\vert \Phi _{k} (n)\rangle$ and 
$\vert \Phi _{l} (n')\rangle$ with $n=n'+1$.

\subsection{Computational details}
We choose to demonstrate the numerical application of the theory
introduced above and in the Appendix (Section \ref{Appendix}) on the Na$_2$ molecule.
In our simulations we consider the ${\rm X}\,^{1}\Sigma{\rm _{g}^{+}}$ ground
and the first excited ${\rm A}\,^{1}\Sigma{\rm _{u}^{+}}$
electronic states of Na$_2$, for which we use the potential energy curves (PEC)
and the transition dipole function of Refs. \cite{Na2_PEC} 
and \cite{Na2_TDM}, respectively.
The field-free rovibrational eigenstates of Na$_2$ on the $V_{\rm X}(R)$ 
and $V_{\rm A}(R)$ PECs are computed using 200 spherical-DVR basis 
function \cite{D2FOPI_PCCP_Szidarovszky2010} with the related grid points
placed in the internuclear coordinate range $(0,10)$ bohr. 
All rovibrational eigenstates with $J<16$ and an energy not exceeding
the zero-point energy of the respective PEC by more than 2000 cm$^{-1}$ were
included into the basis representing the Floquet Hamiltonian of
Eq. (\ref{eq:HF_detailed_2x2}), which was then diagonalized to obtain the light-dressed states.

In all computations the polarization vector of the probe pulse was assumed
to be parallel to the polarization vector of the pump pulse, meaning that the projection of the total angular momentum onto this axis is a conserved quantity during our simulations.

When investigating the effect of the turn-on time of the dressing field (see Sec. \ref{seubsection_effect_of_turn_on_time}), the TDSE was solved using the simple formula $\mathbf{\Psi}(t+dt) = e^{-(i/\hbar)\mathbf{H}(t)dt}\mathbf{\Psi}(t)$. Due to the small size of $\mathbf{H}(t)$ (few thousand by few thousand) the exponential function could be constructed by diagonalizing $\mathbf{H}(t)$ at each time step.

\begin{figure}[t!]
\includegraphics[width=0.95\columnwidth]{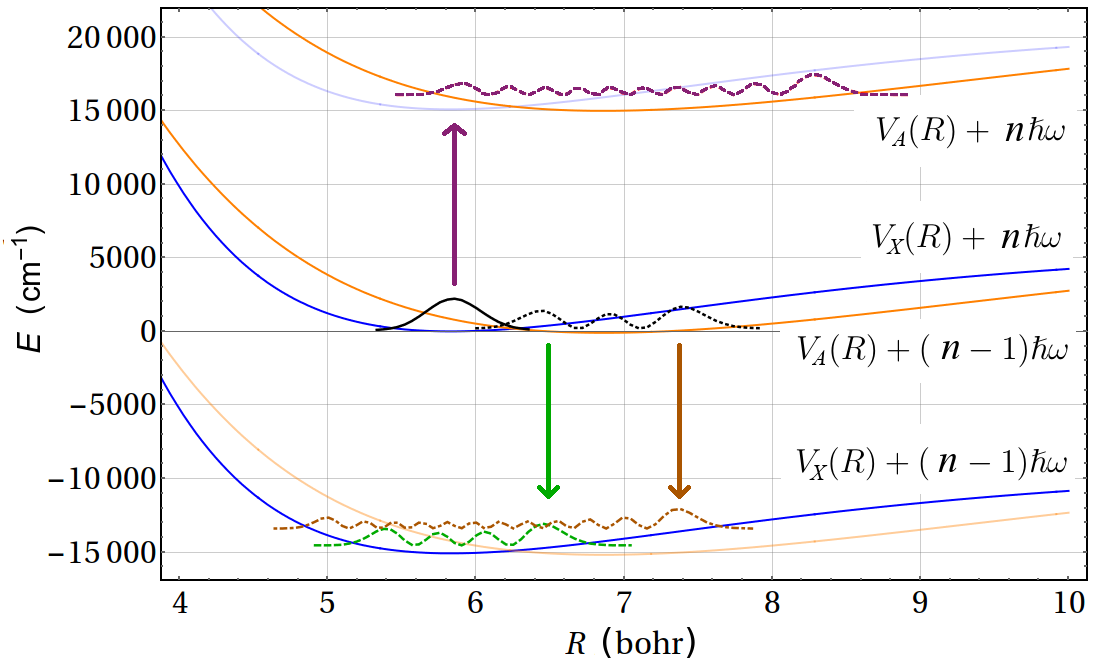}
        
\caption{
Light-dressed diabatic PECs of Na$_{2}$ obtained with a dressing-light
wavelength of $\lambda=657$ nm which is near resonant with the transition between the $\vert {\rm X}$ $0$ $0\rangle$ and $\vert {\rm A}$ $2$ $1\rangle$ states.
The energy scale stands for quasienergy.
Vibrational probability densities are drawn for the
$\vert {\rm X}$ $0$ $0\rangle \vert n \rangle$ (continous black line on the
$V_{X}(R)+n\hslash\omega_1$ PEC), $\vert {\rm X}$ $3$ $0\rangle\vert n-1 \rangle$
(green dashed line on the $V_{X}(R)+(n-1)\hslash\omega_1$ PEC), 
$\vert {\rm X}$ $11$ $0\rangle\vert n-1 \rangle$ (brown dashed line on the 
$V_{X}(R)+(n-1)\hslash\omega_1$ PEC), $\vert {\rm A}$ $2$ $1\rangle\vert n-1 \rangle$ 
(black dotted line on the $V_{A}(R)+(n-1)\hslash\omega_1$ PEC), and 
$\vert {\rm A}$ $9$ $1\rangle\vert n \rangle$ 
(purple dashed line on the $V_{A}(R)+n\hslash\omega_1$ PEC) states.
Upward- and downward pointing vertical arrows represent transitions of
absorption and stimulated emission, respectively.
The two product states with the largest contribution to the light-dressed state
correlating to $\vert {\rm X}$ $0$ $0\rangle$ at $\vert E_1 \vert \rightarrow0$ are
$\vert {\rm X}$ $0$ $0\rangle\vert n \rangle$ and $\vert {\rm A}$ $2$ $1\rangle\vert n-1 \rangle$.\\
}
        \label{fig:PECs}
    \end{figure}

\section{Results and discussion}

\subsection{Interpretation of the light-dressed spectrum}

Before investigating the light-dressed spectra in detail,
it is worth considering their expected structure qualitatively.
Naturally, the light-dressed spectrum strongly depends on the specific
molecule investigated and the properties of the dressing field.
For Na$_2$ the rotational, vibrational, and electronic transition frequencies considered in this work are around the order of 1, 100, and 15\,000 cm$^{-1}$, respectively.   
Figure \ref{fig:PECs} demonstrates the landscape of light-dressed PECs for
the Na$_2$ molecule dressed by a $\lambda=657$ nm wavelength light, which is near resonant with the transition between the $\vert {\rm X}$ $0$ $0\rangle$ and $\vert {\rm A}$ $2$ $1\rangle$ states.
As can be seen in Fig. \ref{fig:PECs}, the manifolds of light-dressed
states, labeled by \textit{n}, are well separated.
Based on Eq. (\ref{eq:Floquet_state_transition_amplitude_2x2_text_body}),
arrows are drawn to indicate the physical origin of possible absorption
and stimulated emission processes induced by the probe pulse.
As seen in Fig.~\ref{fig:PECs}, absorption originates from the first term
in Eq. (\ref{eq:Floquet_state_transition_amplitude_2x2_text_body}),
in which the initial light-dressed state contributes with its X ground
electronic state component
and the final state contributes with
its A excited electronic state component.
On the other hand, stimulated emission originates from the second term
in Eq. (\ref{eq:Floquet_state_transition_amplitude_2x2_text_body}),
in which the initial light-dressed state contributes with its A excited
electronic state component
and the final state contributes with its X ground electronic state component.

\begin{figure}[t!]
    \includegraphics[width=0.90\columnwidth]{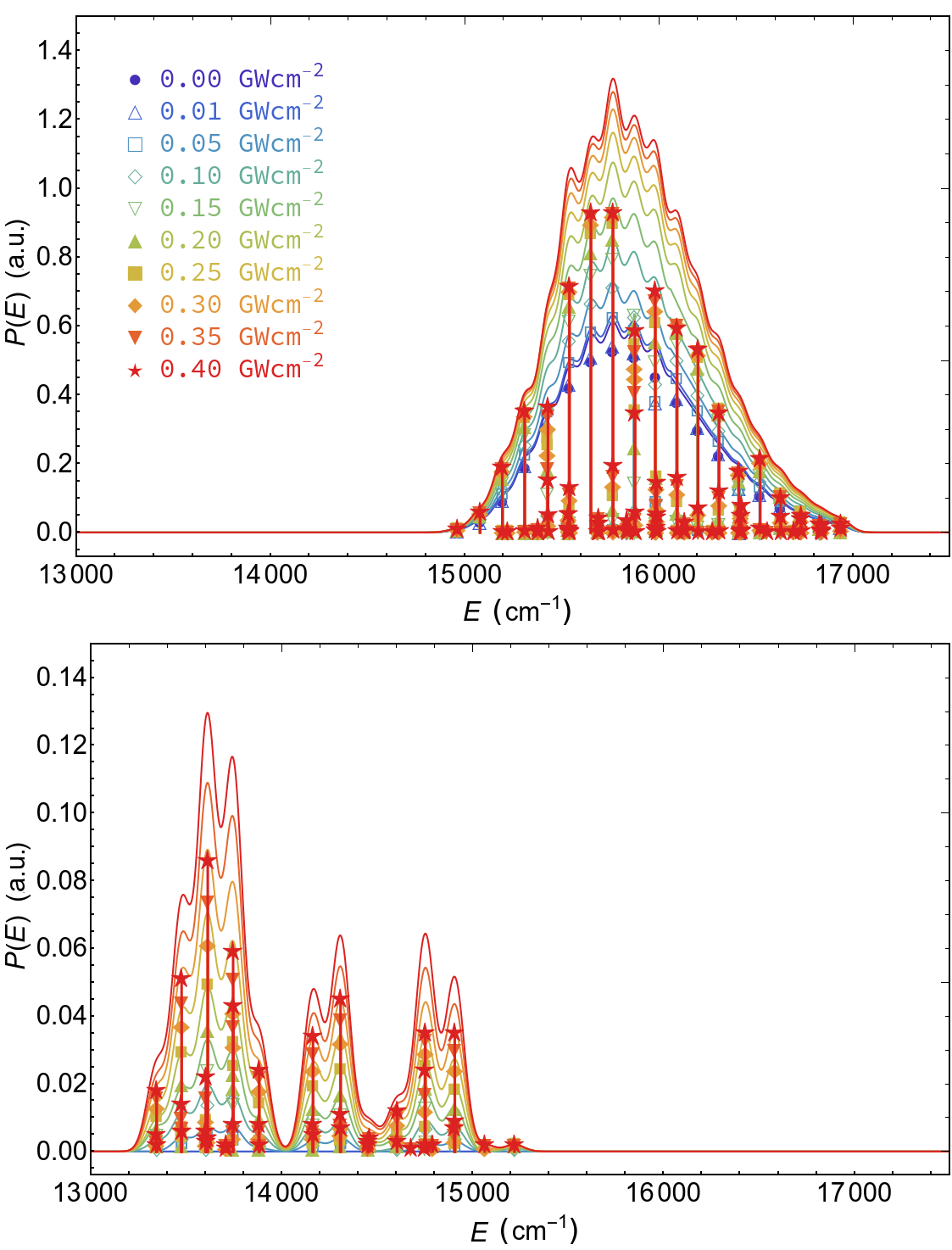}
\caption{
Absorption (upper panel) and stimulated emission (lower panel) spectra
of Na$_2$ dressed with a 657 nm wavelength laser light at 0 K.
The stick spectra were computed using Eq. (\ref{eq:Floquet_state_transition_amplitude_2x2_text_body}) and show transitions from the field-dressed state, which correlates to the $\vert {\rm X}$ $0$ $0\rangle$ rovibronic ground state in the limit of the dressing field intensity going to zero. The envelopes shown are obtained by taking the convolution of the stick spectra with a Gaussian function having a standard deviation of $\sigma=50$ cm$^{-1}$.
}
    \label{fig:FD_spectrum_657nm_0K}
\end{figure}

\subsection{Intensity dependence of the light-dressed spectrum}

Figure \ref{fig:FD_spectrum_657nm_0K} shows the 0~K absorption and
stimulated emission spectra of Na$_2$, when the molecule is dressed
by 657 nm wavelength light fields of different intensity.
Note that there is no special reason for choosing 657 nm to demonstrate the intensity dependence of the light-dressed spectrum, it is merely a convenient choice for which the light-dressed PECs and some relevant states are already depicted in Fig. \ref{fig:PECs}.
    
As seen in Fig. \ref{fig:FD_spectrum_657nm_0K}, with increasing
dressing-field intensity the envelopes of both the absorption and
the stimulated emission spectra increase.
At the limit of zero dressing-field intensity,
the stimulated emission peaks disappear, as expected.
We point out here that the light-dressed states and the
corresponding light-dressed spectra change if the dressing field
wavelength is changed.
Therefore, if dressing light wavelengths different from 657 nm are used,
the envelope of the absorption spectrum might decrease or even show
no monotonic behaviour with increasing dressing light intensity.
    
\begin{figure}[t!]
    \includegraphics[width=0.95\columnwidth]{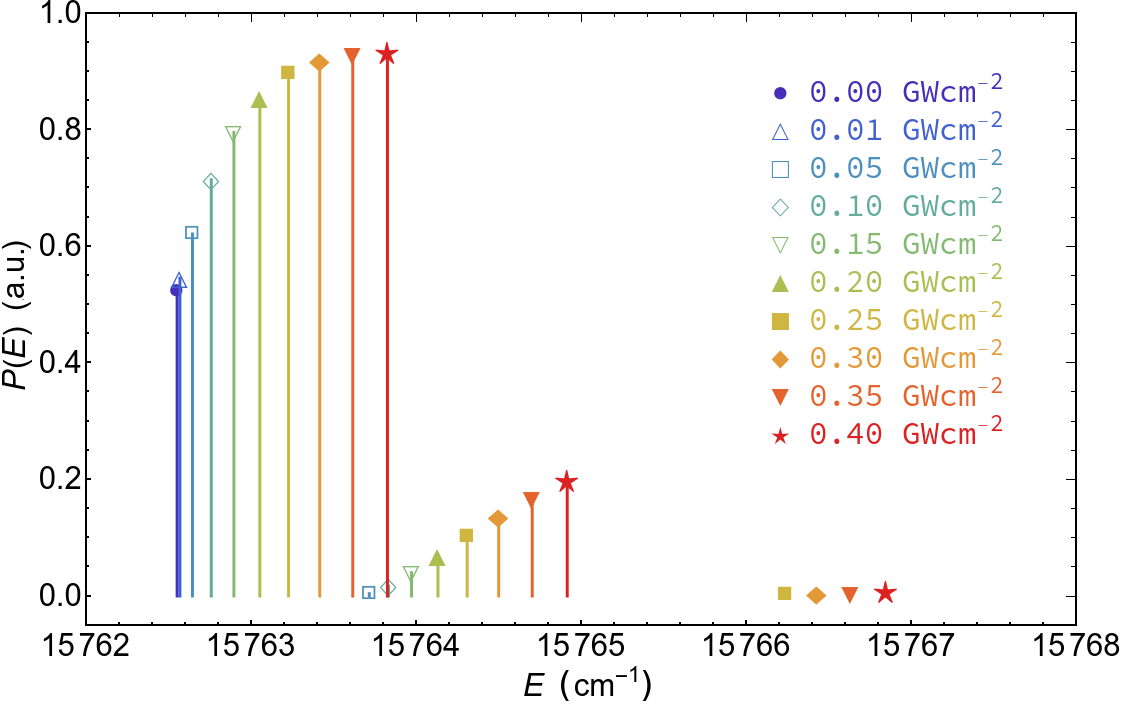}
    \includegraphics[width=0.95\columnwidth]{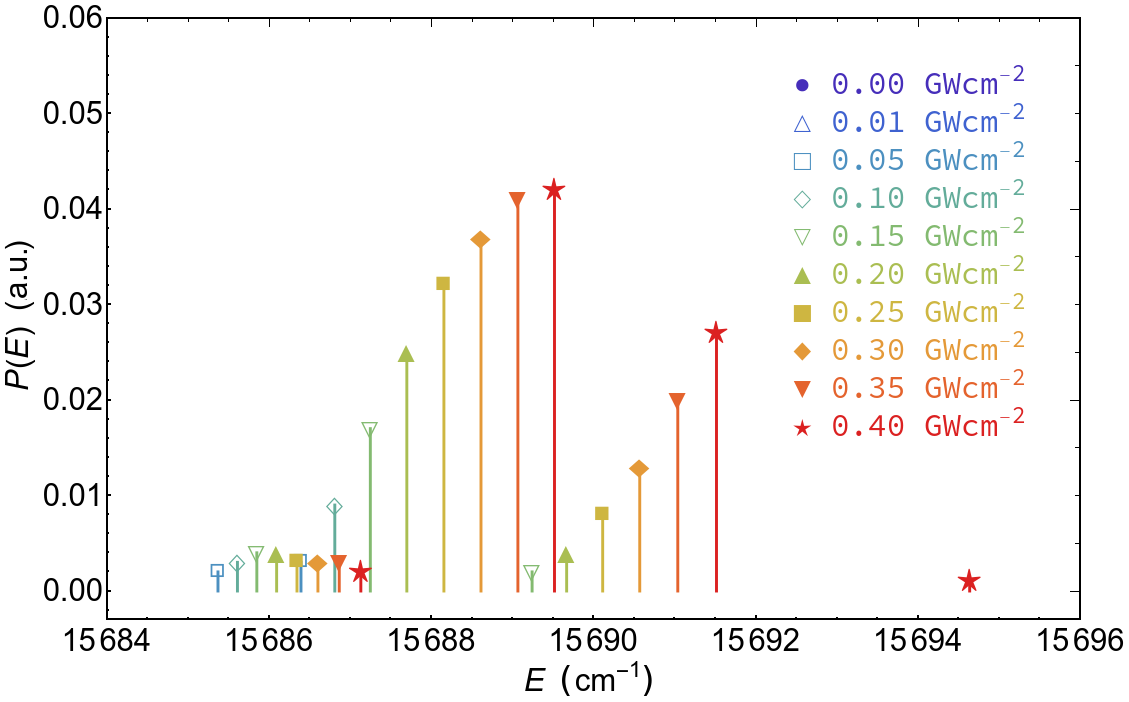}
\caption{
Progression of selected light-dressed absorption peaks of Na$_2$ dressed
with a 657 nm wavelength laser light at 0 K.
The upper panel shows peaks originating from the field-free transition 
$\vert {\rm X}$ $0$ $0\rangle\rightarrow\vert {\rm A}$ $7$ $1\rangle$,
while the lower panel shows peaks which have no field-free counterpart.\\
    }
    \label{fig:peak_progression_abs}
\end{figure}

Inspecting the individual transition lines reveals that introducing
the dressing field leads to the splitting of existing field-free absorption peaks
as well as the appearance of new peaks, as shown in the upper and lower panels
of Fig. \ref{fig:peak_progression_abs}, respectively.  
It is possible to understand and assign the transition peaks of the
light-dressed spectrum based on
the selection rules of transitions
between field-free states and the fact that the
light-dressed states can be described as a superposition of field-free states.

For example, the upper panel of Fig.~\ref{fig:peak_progression_abs} shows
the progression of three peaks, corresponding to transitions from the
initial state (light-dressed state correlating to the field-free ground state)
composed primarily of the $\vert {\rm X}$ $0$ $0\rangle$ state with smaller contributions
from the $\vert {\rm X}$ $0$ $2\rangle$ and $\vert {\rm A}$ $2$ $1\rangle$ states
to light-dressed states
composed primarily of the $\vert {\rm A}$ $7$ $1\rangle$, $\vert {\rm A}$ $7$ $3\rangle$,
and $\vert {\rm A}$ $7$ $5\rangle$ states, with $\vert {\rm X}$ $v$ $J\rangle$-type states
($J$ even) contributing as well. 
These transitions can be interpreted as originating from the field-free transition
$\vert {\rm A}$ $7$ $1\rangle\leftarrow\vert {\rm X}$ $0$ $0\rangle$,
which is split due to the mixing of field-free states through the
light-matter coupling with the dressing field.
Such type of peak splittings are similar in spirit to the
well-known \textit{Autler--Townes effect} \cite{AutlerTownes_original} utilized
in spectroscopy, see, for example,
Refs. \citenum{AutlerTownes_in_spectroscopy_1997,AutlerTownes_in_spectroscopy_2012,AutlerTownes_in_spectroscopy_2017}.
Furthermore,
the upper panel of Fig.~\ref{fig:peak_progression_abs} not only demonstrates
splitting of the peak, during which the sum of individual peak intensities
remain unchanged, but exhibits an overall increase in the peak intensities
when the strength of the dressing field is increased.
In a spectroscopic context the changes in transition peak intensities resulting
from couplings between eigenstates of a zeroth-order Hamiltonian is
usually called \emph{intensity borrowing} \cite{98BuJe}.
    
The lower panel of Fig.~\ref{fig:peak_progression_abs} shows
the progression of three peaks, which do not arise from the splitting of an
existing field-free peak but appear as new peaks.
These transitions occur between the initial state 
(light-dressed state correlating to the field-free ground state) composed
primarily of $\vert {\rm X}$ $0$ $0\rangle$ with smaller contributions from
$\vert {\rm X}$ $0$ $2\rangle$ and $\vert {\rm A}$ $2$ $1\rangle$ and light-dressed states
composed primarily of the $\vert {\rm X}$ $3$ $0\rangle$, $\vert {\rm X}$ $3$ $2\rangle$,
and $\vert {\rm X}$ $3$ $4\rangle$ states. 
Such transitions are forbidden in the limit of zero dressing-light intensity;
however, they become visible as the light-matter coupling with the dressing
field contaminates the $\vert {\rm X}$ $3$ $J\rangle$ states ($J$ even) with
$\vert {\rm A}$ $v$ $1\rangle$-type states, to which $\vert {\rm X}$ $0$ $0\rangle$
has allowed transitions.
The appearance of transition peaks as a result of such a mixing phenomenon
can be understood as an intensity borrowing effect.
    
As to the stimulated emission peaks shown in the lower panel of
Fig.~\ref{fig:FD_spectrum_657nm_0K}, they represent transitions from
the initial state to light-dressed states composed primarily of
vibrationally highly excited $\vert {\rm X}$ $v$ $0\rangle$- and 
$\vert {\rm X}$ $v$ $2\rangle$-type states, with 
$\vert {\rm A}$ $v'$ $J\rangle$-type states ($J$ odd) contributing as well.

\subsubsection{Predicting field-free properties \emph{via} extrapolation}
        
Although light-dressed spectroscopy might provide transition peaks
forbidden in the field-free case, the transition frequencies
between light-dressed states are in general different from the transition
frequencies between field-free states.
If one is interested in obtaining field-free transition frequencies, one might record the light-dressed spectrum at several dressing-field
intensities and extrapolate to the zero intensity limit.
Such a procedure is of course most valuable if the transition is forbidden
in the field-free case.

\begin{figure}[t!]
\includegraphics[width=0.95\columnwidth]{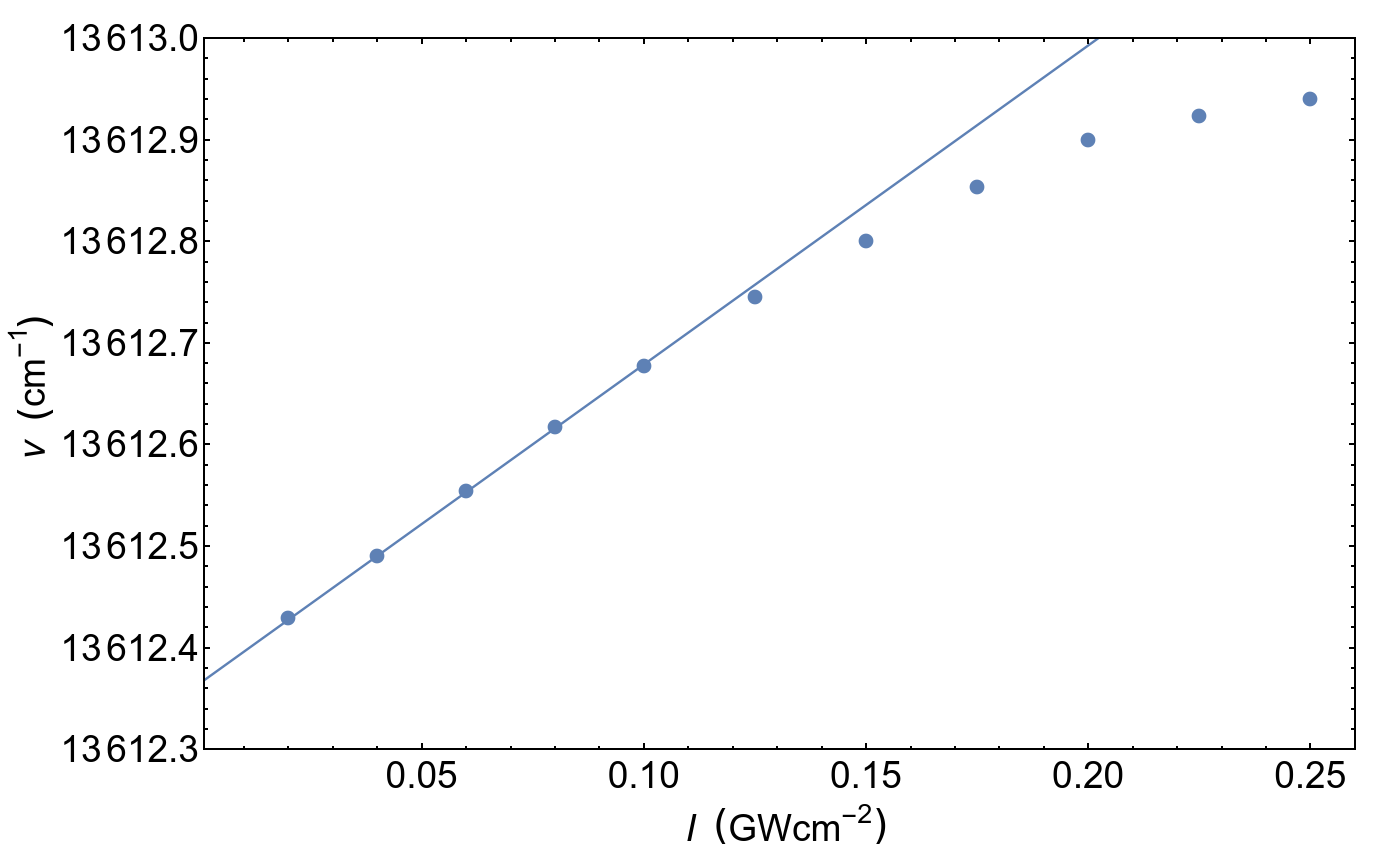}
    \caption{
Position of the stimulated emission peak near 13\,612.5 cm$^{-1}$
as a function of dressing-light intensity for $\lambda=657$ nm.
The straight line plotted was obtained by fitting a linear function
to the first four data points.}
    \label{fig:transition_extrapolation}
\end{figure}

As an example, we examine the stimulated emission peak at around
13\,612.5 cm$^{-1}$, see Fig. \ref{fig:transition_extrapolation}.
The emission peak around 13\,612.5 cm$^{-1}$ represents a transition
in which the initial state is composed primarily of the
$\vert {\rm X}$ $0$ $0 \rangle$ ground state with the
$\vert {\rm X}$ $0$ $2 \rangle$ and
$\vert {\rm A}$ $2$ $1 \rangle$ states contributing as well, and in which
the final state is composed primarily of $\vert {\rm X}$ $11$ $0 \rangle$, 
with some contribution also from $\vert {\rm X}$ $11$ $2 \rangle$ and $\vert {\rm A}$ $17$ $1 \rangle$.
In the limit of zero dressing-light intensity,
the initial and final states correlate to $\vert {\rm X}$ $0$ $0 \rangle$
and $\vert {\rm X}$ $11$ $0 \rangle$, respectively.
Transition between these two field-free states is forbidden; nonetheless,
their accurate transition frequency can be obtained by extrapolating the
light-dressed transition frequency to the limit of zero dressing-light intensity.
As expected and seen in Fig. \ref{fig:transition_extrapolation},
linear extrapolation might be pursued if data points at low dressing-light
intensities are used.
On the other hand, by increasing the dressing-light intensity above a
certain point, the relation between intensity and transition frequency
becomes nonlinear.
As seen in Fig.~\ref{fig:transition_extrapolation}, the value of
the transition frequency extrapolated to zero intensity is 13\,612.37 cm$^{-1}$.
Considering that in this transition the $\vert {\rm X}$ $0$ $0 \rangle$ and
$\vert {\rm X}$ $11$ $0 \rangle$ states belong to the $n$ and $n-1$ Fourier manifolds,
see Fig. \ref{fig:PECs}, the transition frequency between 
$\vert {\rm X}$ $0$ $0 \rangle$ and $\vert {\rm X}$ $11$ $0 \rangle$ can be obtained 
as $(-13\,612.37+15\,220.70)$ $=1608.33$ cm$^{-1}$, 
where $15\,220.70$ cm$^{-1}$ is the photon energy of the dressing light.
The numerical value for the transition wavenumber, obtained as the
difference between the computed field-free eigenenergies of $\vert {\rm X}$ $0$ $0 \rangle$ and
$\vert {\rm X}$ $11$ $0 \rangle$, 
is also $1608.33$ cm$^{-1}$; thus, the extrapolation technique works perfectly.

Of course there are spectroscopic methods already available, such as Laser Induced Disperse Fluoroesence (LIDF) and Stimulated Emission Pumping (SEP) \cite{SEP_and_LIDF_book}, capable of producing data similar to that retrieved from the normally-forbidden transitions measured by our extrapolation scheme. Nonetheless, light-dressed spectroscopy could complement these existing emission-only spectroscopic methods and it is worth noting that the extrapolation scheme can be utilized in both absorption and emission measurements, and varying the dressing-light wavelength could provide control and selectivity over the transitions to be measured, see below.

\subsection{Frequency dependence of the light-dressed spectrum}
    
\begin{figure}[t!]
   \includegraphics[width=0.95\columnwidth]{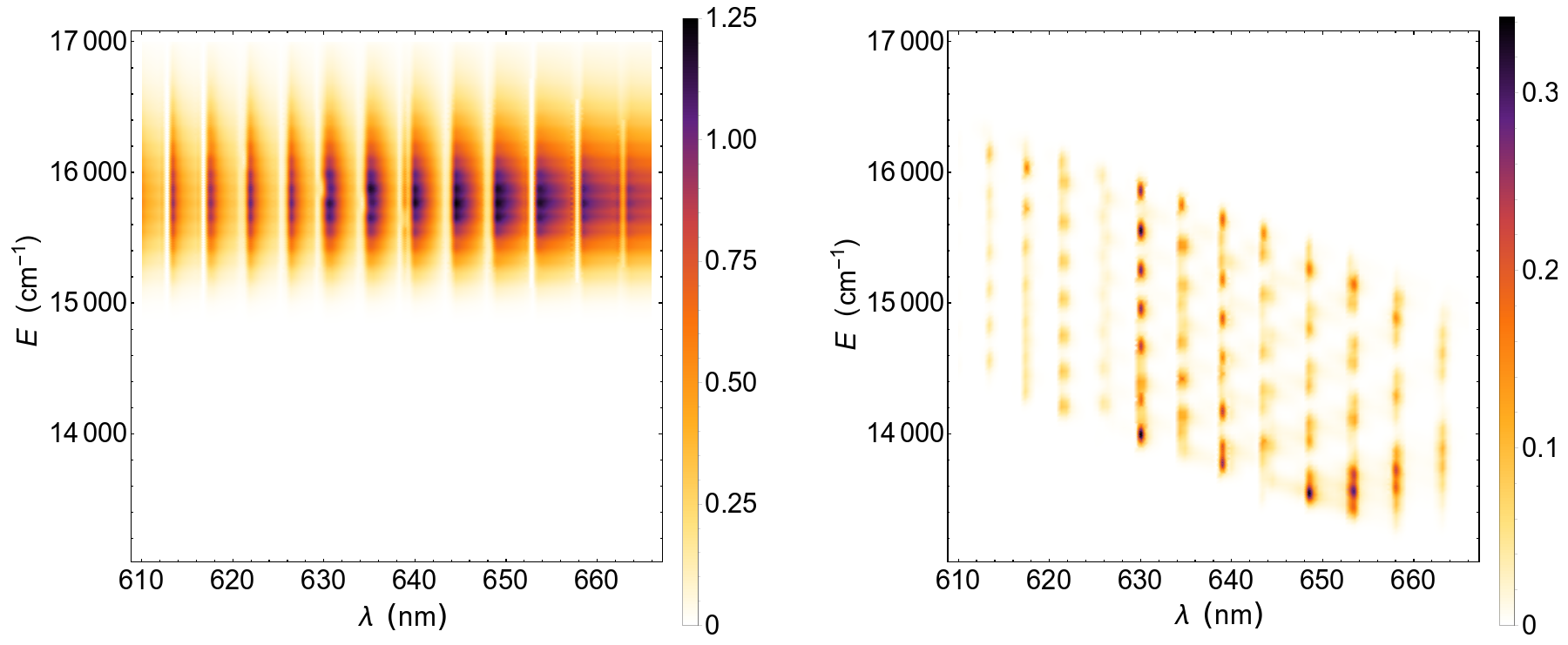}
\caption{
Absorption (left panel) and stimulated emission (right panel) 0 K spectra of
Na$_2$ dressed with an $I=10^8$ W\,cm$^{-2}$ intensity laser light of
different wavelengths.
}
  \label{fig:frequency_sweep_0p01}
\end{figure}

Figure \ref{fig:frequency_sweep_0p01} shows the light-dressed spectrum
of Na$_2$ when dressed by $I=10^8$ W\,cm$^{-2}$ intensity light fields
of different wavelength.
As seen in Fig. \ref{fig:frequency_sweep_0p01},
both the absorption and the stimulated emission spectra vary strongly
with the dressing-light wavelength.
This is expected because with varying dressing-light wavelength
the contribution of different field-free states in the light-dressed states
also vary, leading to varying transition probabilities.
Therefore, by changing the dressing-light wavelength,
one has certain control over which type of transitions appear in the
light-dressed spectrum. 
In the vicinity of dressing light wavelengths which are resonant with a $\vert {\rm A}$ $v$ $1\rangle\leftarrow\vert {\rm X}$ $0$ $0\rangle$-type transition, the absorption spectrum signal decreases (vertical white lines in the left panel of Fig. \ref{fig:frequency_sweep_0p01}) while stimulated emission increases simultaneously. This can be understood as resulting from the increased mixing of field-free states near resonance, \textit{i.e.}, the weight of $\vert {\rm X}$ $0$ $0\rangle$ decreases (leading to a decrease in the absorption), while the weight of $\vert {\rm A}$ $v$ $1\rangle$ increases (leading to an increase in the emission) in the light-dressed state correlating to $\vert {\rm X}$ $0$ $0\rangle$.

Interestingly, due to the values of the Frank--Condon overlaps
between the vibrational states of the X and A electronic states of Na$_2$,
the number of vertical nodes in the stimulated emission spectrum at
different dressing-light wavelengths can in some cases reveal which
$v$ value of the $\vert {\rm A}$ $v$ $J \rangle$-type states contributes
the most to the initial light-dressed state.
For example, using 662~nm dressing light leads to emission lines
whose transition amplitudes primarily originate from 
$\vert {\rm X}$ $v$ $J\pm1\rangle\leftarrow\vert {\rm A}$ $1$ $J\rangle$-type transitions,
while using 657~nm dressing light leads to emission lines whose
transition amplitudes primarily originate from 
$\vert {\rm X}$ $v$ $J\pm1\rangle\leftarrow\vert {\rm A}$ $2$ $J\rangle$-type transitions.

\subsection{Light-dressed spectra at finite temperatures}
    
Up to this point the light-dressed spectra shown correspond to $T=0$ K;
that is, it was assumed that the initial light-dressed state correlates
to the field-free rovibronic ground state of Na$_2$. 
The physical picture behind this assumption is as follows: initially 
the field-free molecules are all in their ground state and these are changed
into dressed states with the adiabatic turn-on of the dressing field.
In a realistic experiment at a finite temperature, however, 
the molecules are not all necessarily in their ground state.
Thus, thermal averaging of the computed spectrum needs to be carried out.
Since it is assumed that thermal averaging occurs only prior but not after
the light-dressing process, the thermal averaging can be done by weighting
transitions with the Boltzmann weights of the field-free states correlating
to the respective initial light-dressed states. 
That is, transitions from each $\vert \Phi _{i} \rangle$ light-dressed state
are considered in the computed spectrum, 
but with all transitions from a given $\vert \Phi _{i} \rangle$ 
light-dressed state weighted by 
    \begin{equation}
        \frac{e^{-E_i^{\rm FF}/kT}}{Q(T)},
        \label{eq:transition_amplitude_between_FD_states_finite_temp}
    \end{equation}
where $Q(T)=\sum_{i}e^{-E_i^{\rm FF}/kT}$ is the rovibronic partition function
of the field-free molecule and $E_i^{\rm FF}$ is the energy of
the field-free rovibronic state to which $\vert \Phi _{i} \rangle$ correlates
in the limit of the dressing light intensity going to zero.
An additional complication at finite temperatures is that one needs to take into
account that the field-free rotational states of a closed-shell diatomic molecule
are characterized not only by $J$, but also by the $m$ quantum number,
which stands for the projection of the rotational angular momentum onto
the chosen space-fixed quantization axis. 
Since a linearly polarized dressing field (and the weak probe pulse 
with identical polarization) can not mix states with different $m$ quantum numbers,
one can simply work in the $m=0$ manifold when $T=0$ K.
For finite-temperature calculations, however, one needs to determine the 
light-dressed states and corresponding transitions for the different $m$ manifolds,
and include them into the spectrum with the appropriate weights shown in 
Eq. (\ref{eq:transition_amplitude_between_FD_states_finite_temp}).
   
\begin{figure}[ht]
        \includegraphics[width=0.95\columnwidth]{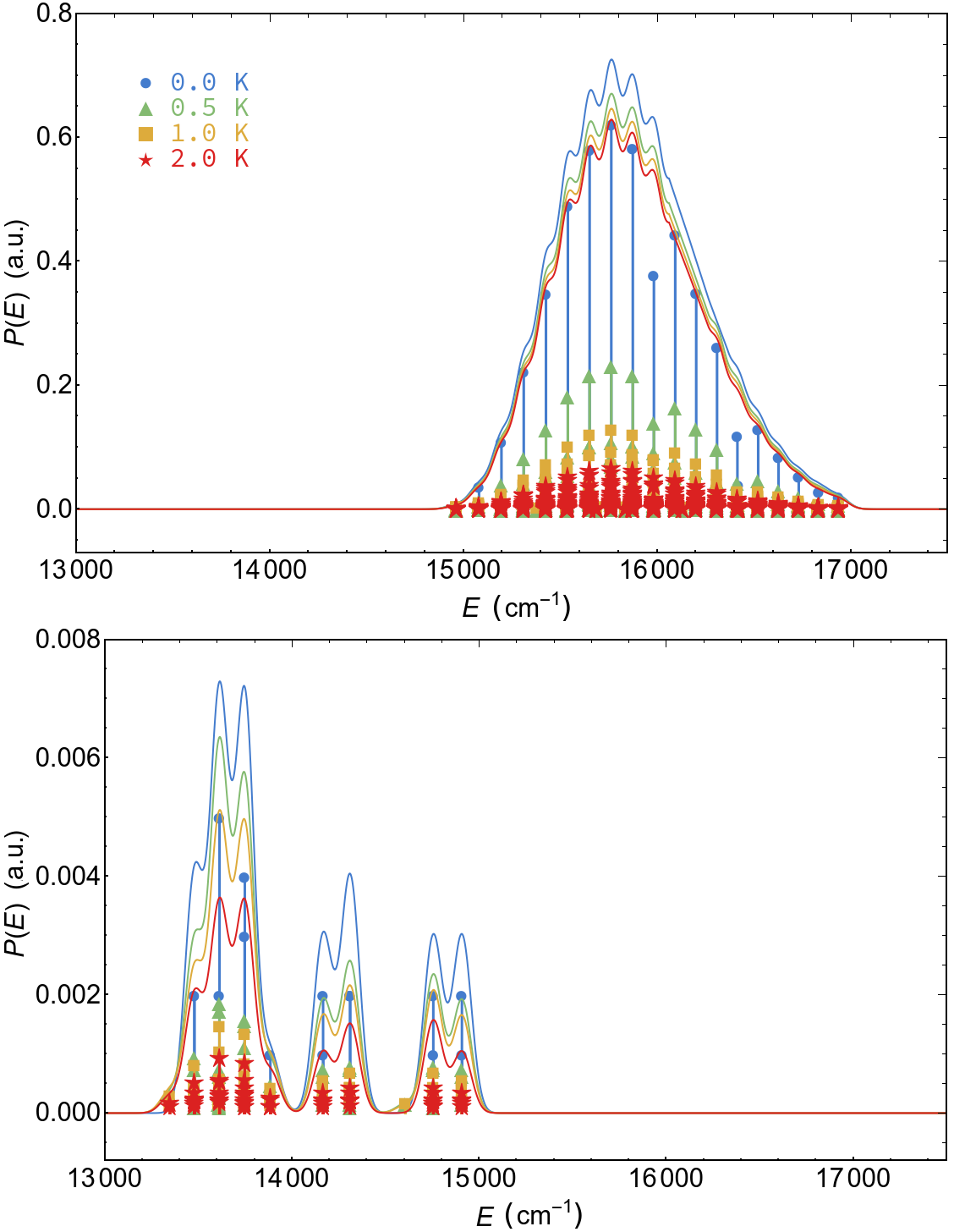}
\caption{Light-dressed absorption (upper panel) and stimulated emission (lower panel)
spectra of Na$_2$ obtained at different temperatures with a 657 nm wavelength dressing
field having 0.05 GW\,cm$^{-2}$ intensity.
        }
        \label{fig:FD_spectrum_657nm_0p05GWcm2}
    \end{figure}

    
Figure \ref{fig:FD_spectrum_657nm_0p05GWcm2} shows the light-dressed spectra of Na$_2$
at different temperatures,, when Na$_2$ is dressed by a 657 nm wavelength light field of
0.05 GW\,cm$^{-2}$ intensity.
The upper panel of Fig. \ref{fig:FD_spectrum_657nm_0p05GWcm2} demonstrates
that although absorption peaks split and the peak height of individual lines decreases
significantly with increasing temperature (due to low-lying rotational states being populated
at finite temperature), the envelope of the spectrum changes to a much smaller extent.
As for stimulated emission, increasing the temperature seems to have a more pronounced
effect on the spectrum envelope than in the case of absorption, see the lower panel of
Fig.~\ref{fig:FD_spectrum_657nm_0p05GWcm2}.
    
\begin{figure}[ht]
        \includegraphics[width=0.95\columnwidth]{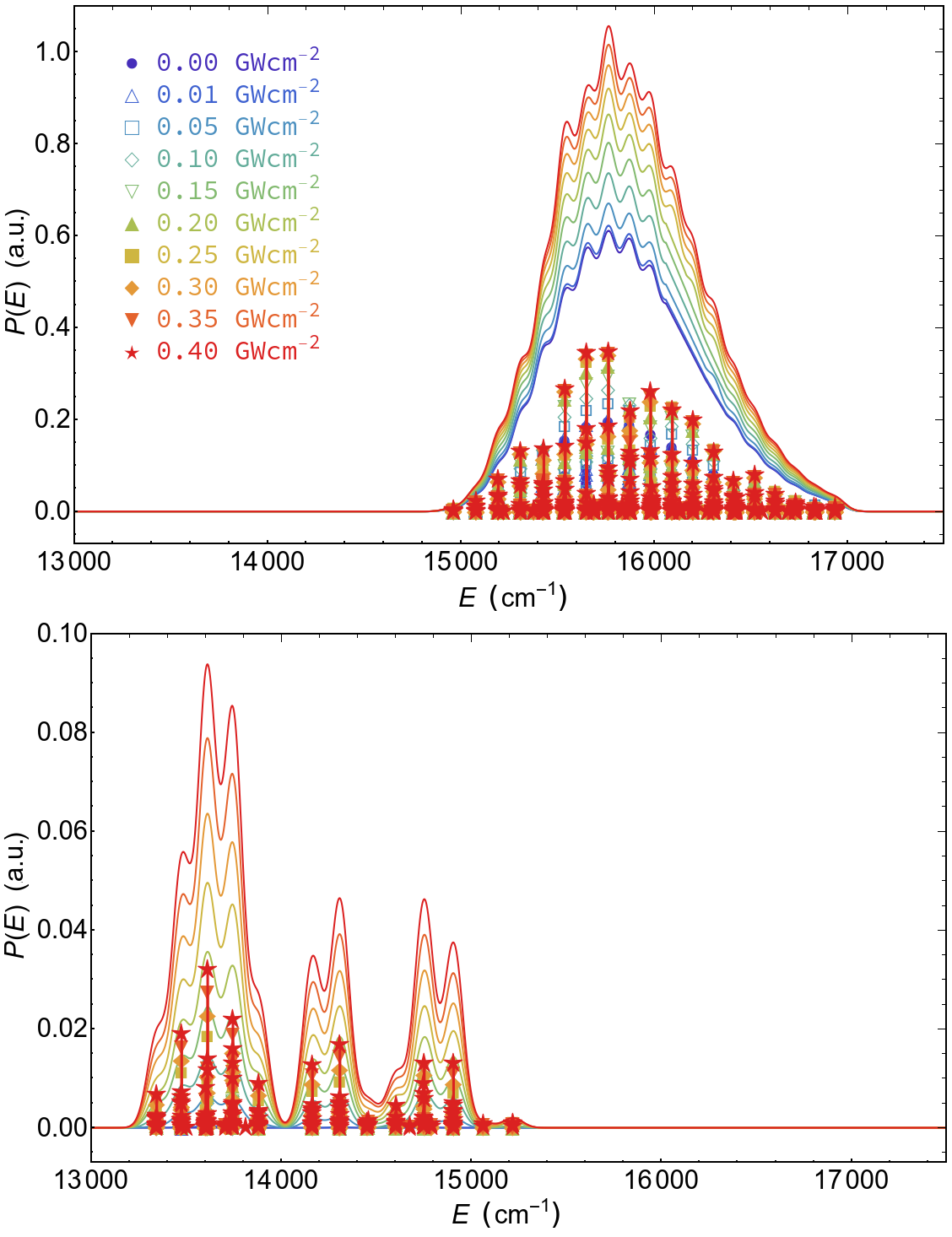}
        
        \caption{
        Light-dressed absorption (upper panel) and stimulated emission (lower panel)
				spectra of Na$_2$ obtained at 0.5~K with 657~nm wavelength dressing fields
				having different intensities.
        }
        \label{fig:FD_spectrum_657nm_0p5K}
    \end{figure}

Figure \ref{fig:FD_spectrum_657nm_0p5K} shows the same dressing-light intensity
dependence of the light-dressed spectra as in Fig. \ref{fig:FD_spectrum_657nm_0K},
but at $T=0.5$ K.
The same conclusions apply as for Fig. \ref{fig:FD_spectrum_657nm_0p05GWcm2};
the number of individual lines and the peak amplitudes are much more affected
by temperature than the spectrum envelopes.

\subsection{Effects of the dressing-field turn-on time on the light-dressed states}
\label{seubsection_effect_of_turn_on_time}

Up to this point it was assumed that the dressing field is turned on adiabatically.
This resulted in the fact that starting from an initial field-free state the
generated light-dressed wave function is composed of a single light-dressed state,
correlating to the initial field-free state.
If the dressing field is not turned on adiabatically, then the generated
light-dressed wave function becomes a superposition of light-dressed states,
with the coefficients depending on the turn-on time. Interested readers can find further information and in-depth investigations on the temporal evolution of light-dressed states in the Floquet formalism, for example, in Refs. \cite{AdiabaticFloquet2_Gurin_AdvChemPhys_2003,AdiabaticFloquet_Lefebvre_PRA_2013,AdiabaticFloquet3_Leclerc_PRA_2016}.

\begin{figure}[ht]
        \includegraphics[width=0.90\columnwidth]{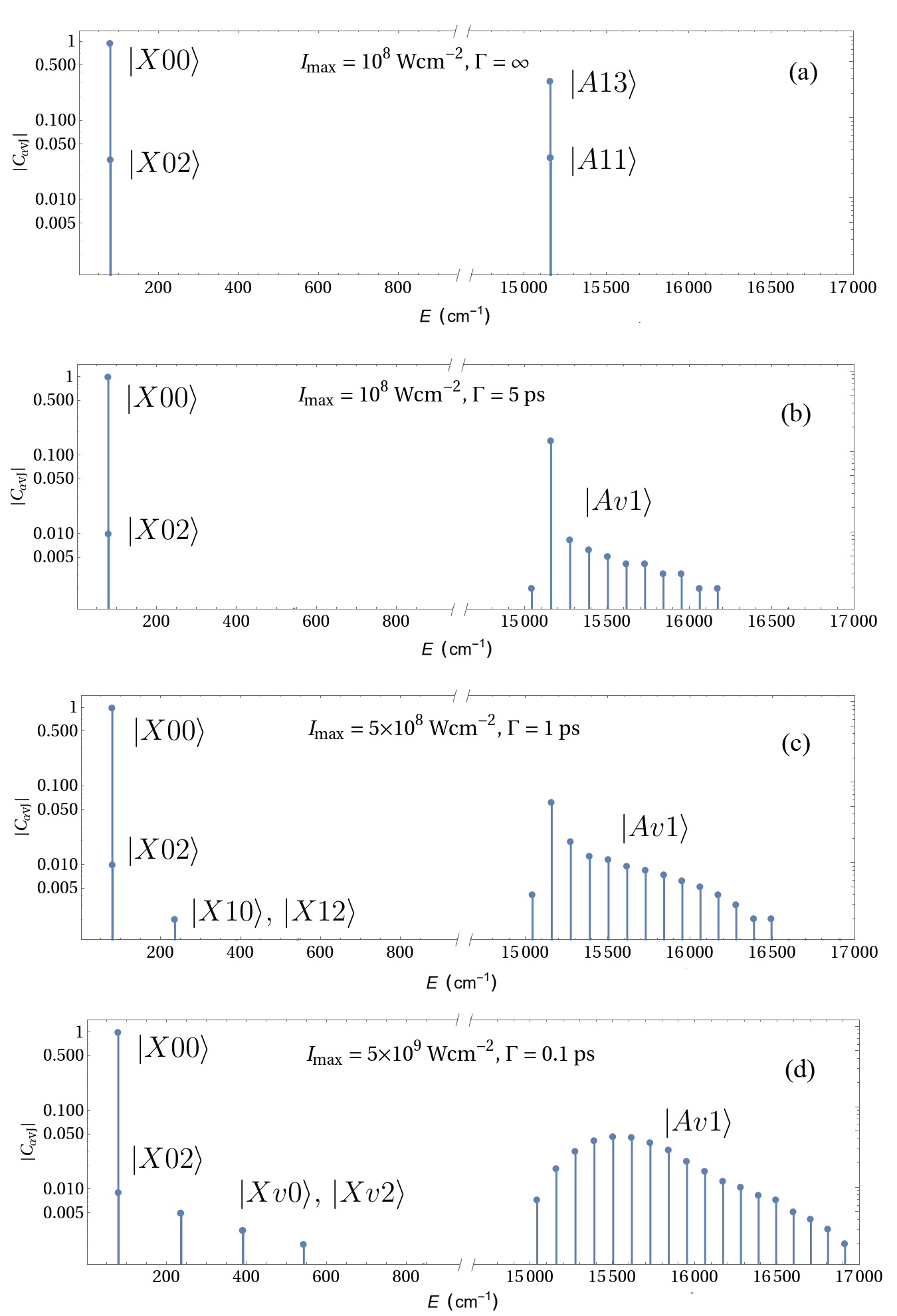}
        
        \caption{
        Population of the different field-free eigenstates in the wave function for $\lambda=663$ nm wavelength dressing-fields of different turn-on time and intensity.
        }
        \label{fig:TD_dynamics}
    \end{figure}

Figure \ref{fig:TD_dynamics} demonstrates the population of the different
field-free eigenstates in the wave function for dressing-fields of different
turn-on time ($\mathit{\Gamma}$) and intensity.
The functional form of the dressing light was assumed to be 
$\bm{E}_1(t) =\bm{0}$ for $t<0$,
$\bm{E}_1(t) = \bm{E}_{\rm max}{\rm sin}(\hbar \omega_1 t){\rm sin}^2(\pi t / \mathit{\Gamma})$
for $0<t<\mathit{\Gamma}/2$, and $\bm{E}_1(t) = \bm{E}_{\rm max}$ for $\mathit{\Gamma}/2<t$.
The populations shown in Fig. \ref{fig:TD_dynamics} were computed for 
$t=\mathit{\Gamma}/2$ by solving the TDSE directly.
    
Panels (b-d) of Fig.~\ref{fig:TD_dynamics} demonstrate that when the
turn-on time ($\mathit{\Gamma}$) is shorter than the characteristic timescale of
molecular rotations (based on the 
$\vert {\rm X}$ $0$ $0\rangle \leftrightarrow \vert {\rm X}$ $0$ $1\rangle$ transition of
Na$_2$ this is around 100 ps), then the degree of rotational excitation is reduced
and in fact in the excited electronic state it is limited to that required
by the optical selection rules ($\Delta J = \pm 1$).
Panels (c) and (d) of Fig. \ref{fig:TD_dynamics} also demonstrate that
as the turn-on time of the dressing light becomes shorter in the time domain,
it expands in the frequency domain, giving rise to significant populations in $\vert {\rm X}~v~0 \rangle$- and
$\vert {\rm A}~v~1 \rangle$-type field-free eigenstates in a wider energy range.
    
\begin{figure}[ht]
        \includegraphics[width=0.90\columnwidth]{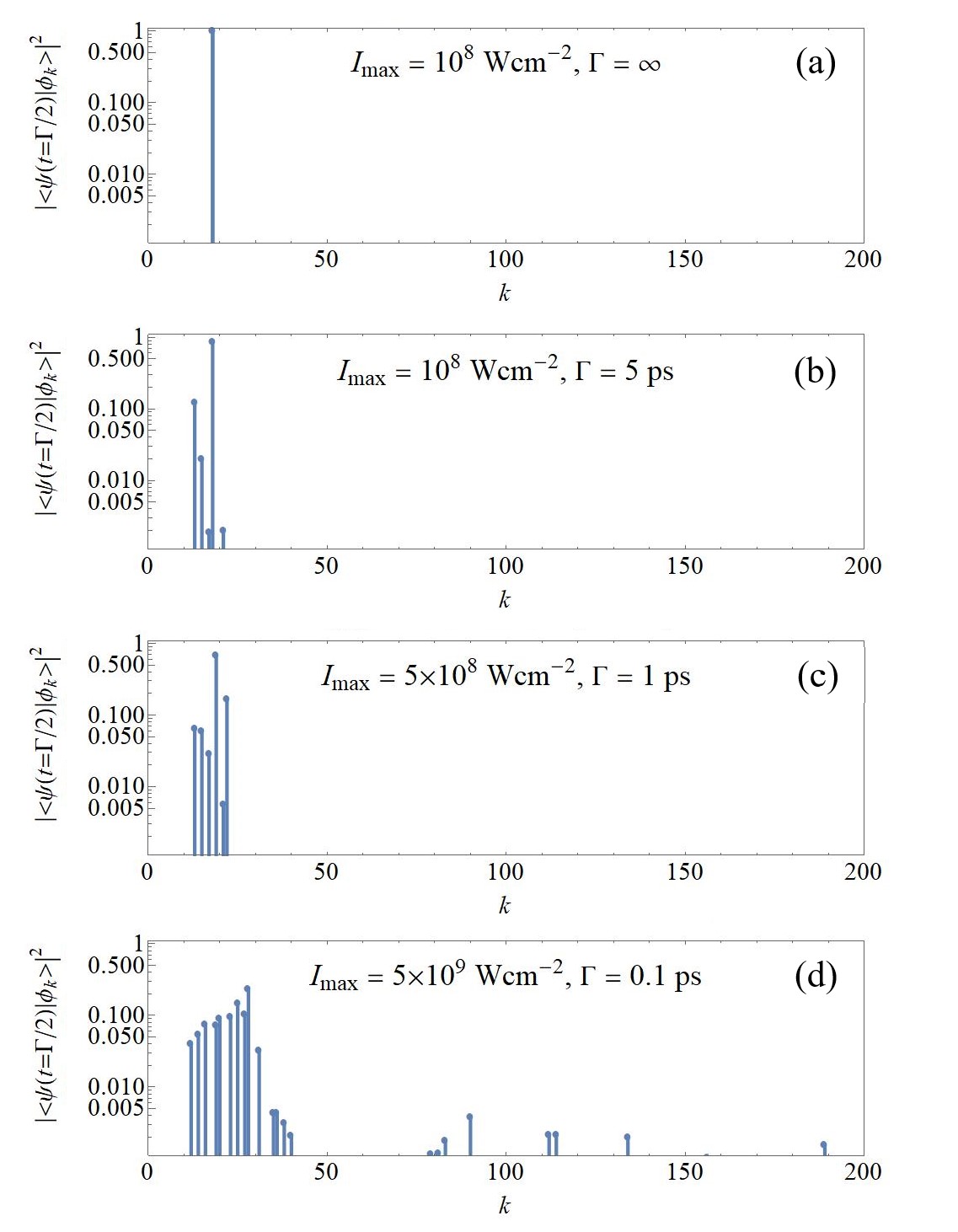}
        
        \caption{
        Population of the different light-dressed states in the wave function for $\lambda=663$ nm wavelength dressing-fields of different turn-on time and intensity.}
        \label{fig:TD_dynamics_FD_overlap}
    \end{figure}

After the dressing field reaches its peak intensity at $t=\mathit{\Gamma}/2$
(and is kept constant thereafter) the wave function can be expanded as a 
superposition of the light-dressed states, and the light-dressed spectrum
can be calculated using Eq. (\ref{eq:transition_amplitude_Floquet_general}).
Figure \ref{fig:TD_dynamics_FD_overlap} depicts the population of the different light-dressed states in the wave function for dressing-fields of different turn-on time ($\mathit{\Gamma}$) and intensity. Fig. \ref{fig:TD_dynamics_FD_overlap} demonstrates that by decreasing the turn-on time of the dressing field the initial wave function becomes a more and more pronounced mixture of light-dressed states. Nonetheless,
if the probe pulse is long enough to average out interferences in the
transition probability, then the light-dressed spectrum can be generated
as a simple weighted sum of the spectra of individual light-dressed states.
    
To summarize, by changing the turn-on time of the dressing field,
the initial wave function, and thus the transition peaks observed in the
light-dressed spectrum can be influenced or controlled, although this
might complicate the interpretation of the spectrum significantly. 
    
\section{Summary and conclusions}

In this work we presented a theoretical method to compute the light-dressed spectra of molecules dressed by medium-intensity classical light fields.
Such \textit{light-dressed spectroscopy} is not only of fundamental interest, but can also be a useful practical tool for providing valuable spectroscopic data and related insight for both light-dressed and field-free molecular systems. The approach is based on Floquet theory for deriving the light-dressed states, which are expanded in the basis of field-free molecular eigenstates. Once the light-dressed states are determined, transition amplitudes between them, as recorded by a weak probe pulse, are computed with perturbation theory.

Numerical applications are demonstrated for the homonuclear diatomic molecule Na$_2$, for which the general formulae can be simplified and the physical processes leading to its rovibronic light-dressed spectra can be understood.

It was found that with respect to the field-free spectrum, the light dressing process leads to the splitting of existing peaks as well as the appearance of new peaks, forbidden in the field-free case. Processes similar in spirit to the well-known \emph{Autler--Townes} and \emph{intensity borrowing} effects of molecular spectroscopy could be identified.

The dependence of the light-dressed rovibronic spectrum of Na$_2$ on the dressing light intensity and the dressing light wavelength was also investigated. It was found that by manupulating the intensity and the wavelength of the dressing light one can influence and to some extent control the peaks appearing in the light-dressed spectra.

Implication to determine the frequencies of forbidden field-free transitions by extrapolating light-dressed transition frequencies to the limit of zero dressing light intensity was also made and the extrapolation scheme seemed to produce excellent results.

Finite temperature calculations, assuming different initial temperatures of the molecular ensemble prior to the dressing process revealed that the individual field-dressed spectral peaks are much more sensitive to the initial temperature than their envelope, similar to the field-free case.

Finally, it was shown that by changing the turn-on time of the dressing field from the adiabatic limit to shorter times, the initial wave function of the light-dressed system can be modified significantly, which is likely to result in changes in the light-dressed spectrum as well.

\section{Acknowledgement}
This research was supported by the EU-funded Hungarian grant EFOP-3.6.2-16-2017-00005.
The authors are grateful to NKFIH for support (Grant No. PD124623 and K119658).

\setcounter{equation}{0}
\renewcommand\theequation{A\arabic{equation}}

\section{Appendix}
\label{Appendix}

\subsection{Determination of light-dressed states} \label{Appendix_A}

\subsubsection{General considerations} \label{Appendix_A_1}
In this section we summarize some aspects of the Floquet
approach \cite{Floquet,Shirley_Floquet_2,Chu1985_Floquet_3} used to compute
the light-dressed states generated by a $\hat{W}_1 (t)$ interaction between
the molecule and the dressing field.
For a Hamiltonian periodic in time, such as
\begin{equation}
    \hat{H}_{\rm d}(t)=\hat{H}_{\rm mol} + \hat{W}_1 (t),
    \label{eq:Hd_Hamiltonian}
\end{equation}
\begin{equation}
    \hat{H}_{\rm d}(t+T)=\hat{H}_{\rm d}(t),
\end{equation}
where $ T = 2\pi / \omega _1$, the time-dependent Schr\"odinger equation (TDSE)
\begin{equation}
    i \hbar \partial _t \vert \psi (t) \rangle = \hat{H}_{\rm d}(t) \vert \psi (t) \rangle
    \label{eq:Hd_TDSE}
\end{equation}
has the general solution of the form
\begin{equation}
    \vert \psi (t) \rangle = \sum_{k} c_k e^{- \frac{i}{\hbar} \varepsilon _k t} \vert \Phi _k (t) \rangle,
    \label{eq:Floquet_solution_general}
\end{equation}
where $\varepsilon _k$ are the so-called quasienergies and the $\vert \Phi _k (t) \rangle$ Floquet states (also termed light-dressed states in our work) satisfy
\begin{equation}
    \vert \Phi _k (t+T) \rangle = \vert \Phi _k (t) \rangle,
\end{equation}
and
\begin{equation}
    (\hat{H}_{\rm d}(t) - i \hbar \partial _t) \vert \Phi _k (t) \rangle = \hat{H}_{\rm F}(t) \vert \Phi _k (t) \rangle = \varepsilon _k \vert \Phi _k (t) \rangle.
    \label{eq:Floquet_state_eigenvalue_problem}
\end{equation}
As can be verified using Eq. (\ref{eq:Floquet_state_eigenvalue_problem}), if $\varepsilon_k$ is a quasienergy, then $\varepsilon_k + \hbar m \omega_1$ is also a quasienergy with a corresponding Floquet state $e^{i m \omega_1 t} \vert \Phi _k (t) \rangle$. However, it is clear from Eq. (\ref{eq:Floquet_solution_general}) that such a shifted quasienergy does not represent a new physical state, because $\varepsilon_k + \hbar m \omega_1$ with $e^{i m \omega_1 t} \vert \Phi _k (t) \rangle$ give the same contribution to the wave function as $\varepsilon_k$ with $\vert \Phi _k (t) \rangle$. 

Because $\vert \Phi _k (t) \rangle$ are periodic in time, they can be expanded as a Fourier series
\begin{equation}
    \vert \Phi _{k} (t) \rangle = \sum_{n} \vert \varphi _{kn} \rangle e^{in\omega_1 t}.
    \label{eq:Floquet_state_Fourier_series}
\end{equation}
Combining Eqs. (\ref{eq:Floquet_state_eigenvalue_problem}) and (\ref{eq:Floquet_state_Fourier_series}) and assuming $\hat{W}_1 (t) = - \mathbf{E}_1 \hat{\boldsymbol{\mu}} {\rm cos}(\omega _1 t)= - \frac{1}{2} \mathbf{E}_1 \hat{\boldsymbol{\mu}} (e^{i \omega _1 t}+e^{- i \omega _1 t})$ gives
\begin{equation}
    (\hat{H}_{\rm mol} + \hbar n \omega_1) \sum_{n} \vert \varphi _{kn} \rangle e^{in\omega_1 t} - \frac{1}{2} \mathbf{E}_1 \hat{\boldsymbol{\mu}} \sum_{n} \vert \varphi _{kn} \rangle (e^{i(n+1)\omega_1 t} + e^{i(n-1)\omega_1 t})  = \varepsilon _k \sum_{n} \vert \varphi _{kn} \rangle e^{in\omega_1 t}.
    \label{eq:Floquet_state_eigenvalue_problem_2}
\end{equation}
Multiplying Eq. (\ref{eq:Floquet_state_eigenvalue_problem_2}) with $\frac{1}{T} e^{- i m \omega_1 t}$ and integrating on the time period $T$ leads to
\begin{equation}
    (\hat{H}_{\rm mol} + \hbar m \omega_1) \vert \varphi _{km} \rangle - \frac{1}{2} \mathbf{E}_1 \hat{\boldsymbol{\mu}} \big ( \vert \varphi _{k,m-1} \rangle + \vert \varphi _{k,m+1} \rangle \big ) = \varepsilon _k \vert \varphi _{km} \rangle.
    \label{eq:Floquet_state_eigenvalue_problem_3}
\end{equation}
The Fourier components $\vert \varphi _{km} \rangle$ can further be expressed as a linear combination of rovibronic molecular states
\begin{equation}
    \vert \varphi _{km} \rangle = \sum_{\alpha,v,J} C^{(k)}_{m,\alpha v J} \vert \alpha v J\rangle,
\end{equation}
where $\alpha,v$ and $J$ represent electronic, vibrational and rotational quantum numbers, respectively. Using such an expansion, Eq. (\ref{eq:Floquet_state_eigenvalue_problem_3}) can be turned into the matrix eigenvalue problem
\begin{equation}
    \sum_{n,\alpha,v,J} (\mathbf{H}_{\rm F})_{m\alpha'v'J',n\alpha vJ}  C^{(k)}_{n,\alpha v J} = \varepsilon _k  C^{(k)}_{m,\alpha' v' J'},
    \label{eq:Floquet_state_eigenvalue_problem_4}
\end{equation}
where
\begin{equation}
    \begin{split}
        (\mathbf{H}_{\rm F})_{m\alpha'v'J',n\alpha vJ}= \\
        \big ( \langle \alpha' v' J'\vert \hat{H}_{\rm mol} \vert \alpha v J \rangle + \hbar m \omega_1 \delta_{\alpha \alpha'} \delta_{vv'} \delta_{JJ'} \big ) \delta_{nm} - \frac{1}{2} \langle \alpha' v' J'\vert \mathbf{E}_1 \hat{\boldsymbol{\mu}} \vert \alpha v J \rangle \big ( \delta_{n,m-1} + \delta_{n,m+1} \big ).
        \label{eq:HF_eq}
    \end{split}
\end{equation}
The pictorial representation of $\mathbf{H}_{\rm F}$ reads
\begin{equation}
    \mathbf{H}_{\rm F}=
     \begin{bmatrix}
     \ddots & \vdots & \vdots & \vdots & \vdots & \vdots & \vdots & \reflectbox{$\ddots$} \\
     \cdots & \mathbf{H}_{\rm A} + \hbar \omega_1 \mathbf{I} & \bm{\lambda} & \mathbf{g}_{\rm AA} & \mathbf{g}_{\rm AX} & 0 & 0 & \cdots \\
     \cdots & \bm{\lambda}^\dagger & \mathbf{H}_{\rm X} + \hbar \omega_1 \mathbf{I} & \mathbf{g}_{\rm XA} & \mathbf{g}_{\rm XX} & 0 & 0 & \cdots \\
     \cdots & \mathbf{g}^\dagger_{\rm AA} & \mathbf{g}^\dagger_{\rm XA} & \mathbf{H}_{\rm A} & \bm{\lambda} & \mathbf{g}_{\rm AA} & \mathbf{g}_{\rm AX} & \cdots \\
     \cdots & \mathbf{g}^\dagger_{\rm AX} & \mathbf{g}^\dagger_{\rm XX} & \bm{\lambda}^\dagger & \mathbf{H}_{\rm X} & \mathbf{g}_{\rm XA} & \mathbf{g}_{\rm XX} & \cdots\\
     \cdots & 0 & 0 & \mathbf{g}^\dagger_{\rm AA} & \mathbf{g}^\dagger_{\rm XA} & \mathbf{H}_{\rm A} - \hbar \omega_1 \mathbf{I} & \bm{\lambda}& \cdots\\
     \cdots & 0 & 0 & \mathbf{g}^\dagger_{\rm AX} & \mathbf{g}^\dagger_{\rm XX} & \bm{\lambda}^\dagger & \mathbf{H}_{\rm X} - \hbar \omega_1 \mathbf{I} & \cdots\\
     \reflectbox{$\ddots$} & \vdots & \vdots & \vdots & \vdots & \vdots & \vdots & \ddots
     \end{bmatrix},\label{eq:HF_detailed}
\end{equation}
where different matrix elements represent different values of the $\alpha$ electronic and $n$ Fourier indices of Eq. (\ref{eq:HF_eq}) (for the sake of simplicity, we assumed only two electronic states, labeled X and A), and each matrix element in Eq. (\ref{eq:HF_detailed}) is itself a matrix representation of different operators in the space of rovibrational states, \textit{i.e.},
\begin{equation}
    ( \mathbf{H}_{\rm A} )_{v'J',vJ}= \langle {\rm A} v' J'\vert \hat{H}_{\rm mol} \vert {\rm A} v J \rangle
\end{equation}
\begin{equation}
    ( \mathbf{H}_{\rm X} )_{v'J',vJ}= \langle {\rm X} v' J'\vert \hat{H}_{\rm mol} \vert {\rm X} v J \rangle
\end{equation}
\begin{equation}
    ( \bm{\lambda} )_{v'J',vJ}= \langle {\rm A} v' J'\vert \hat{H}_{\rm mol} \vert {\rm X} v J \rangle
\end{equation}
\begin{equation}
    ( \mathbf{g}_{\alpha \beta} )_{v'J',vJ}= - \frac{1}{2} \langle \alpha v' J'\vert \mathbf{E}_1 \hat{\boldsymbol{\mu}} \vert \beta v J \rangle
\end{equation}
and $\mathbf{I}$ is the identity matrix. $\mathbf{H}_{\rm A}$ and $\mathbf{H}_{\rm X}$ can be thought of as the rovibrational Hamiltonians in the adiabatic electronic states A and X, respectively, while $\bm{\lambda}$ accounts for intrinsic nonadiabatic couplings between the X and A electronic states. $\mathbf{g}_{\alpha \beta}$ naturally represents the coupling induced by the dressing field.

\subsubsection{Simplifying assumptions} \label{Appendix_A_2}

In practical applications Eq. (\ref{eq:HF_detailed}) can often be simplified. (1) If intrinsic nonadiabatic couplings can be neglected in the system under investigation, then $\bm{\lambda}=\boldsymbol{0}$. (2) If the molecule has no permanent dipole, then $\mathbf{g}_{\rm XX}=\mathbf{g}_{\rm AA}=\boldsymbol{0}$. (3) Finally, if $\hbar \omega_1$ is resonant with the electronic excitation between the states X and A, then nonresonant coupling terms can be neglected up to moderate field strengths 
\cite{Halasz2012}. This leaves only those $\mathbf{g}_{\alpha \beta}$ matrices nonzero which connect $\mathbf{H}_{\rm X} + n \hbar \omega_1 \mathbf{I}$ and $\mathbf{H}_{\rm A} + (n-1) \hbar \omega_1 \mathbf{I}$ type elements. With the above three simplifications $\mathbf{H}_{\rm F}$ becomes block diagonal with each two-by-two block being identical up to a constant shift. The block labeled with the Fourier index \textit{n} reads
\begin{equation}
    \mathbf{H}_{\rm F}^{2\times2}(n)=
     \begin{bmatrix}
       \mathbf{H}_{\rm X} + n \hbar \omega_1 & \mathbf{g}_{\rm XA} \\
     \mathbf{g}^\dagger_{\rm XA} & \mathbf{H}_{\rm A} + (n-1) \hbar \omega_1 \mathbf{I}
     \end{bmatrix}.\label{eq:HF_detailed_2x2}
\end{equation}
Therefore, instead of solving the general case of Eq. (\ref{eq:Floquet_state_eigenvalue_problem_3}), it becomes sufficient to solve the eigenvalue problem for $\mathbf{H}_{\rm F}^{2\times2}(n)$ in order to obtain the light-dressed states and corresponding quasienergies.

\subsection{Temporal evolution of a light-dressed system} \label{Appendix_B}

In the representation utilized in Eq. (\ref{eq:HF_eq}) the $\mathbf{H}_{\rm F}$ matrix of the Floquet Hamiltonian and the $C^{(k)}_{n,\alpha v J}$ expansion coefficients of the Floquet states are time independent. Based on this representation and Eqs. (\ref{eq:Floquet_solution_general}) and (\ref{eq:Floquet_state_eigenvalue_problem}), the temporal evolution of a light-dressed system can be expressed as \cite{Shirley_Floquet_2}
\begin{equation}
    \boldsymbol{\Psi} (t) = \sum_{k} c_k e^{- \frac{i}{\hbar} \mathbf{H}_{\rm F} t} \boldsymbol{\Phi}_k = e^{- \frac{i}{\hbar} \mathbf{H}_{\rm F} t} \sum_{k} c_k \boldsymbol{\Phi}_k = e^{- \frac{i}{\hbar} \mathbf{H}_{\rm F} t} \boldsymbol{\Psi} (t=0),
    \label{eq:Floquet_temporal_evolution}
\end{equation}
which is formally equivalent to the temporal evolution of a system with a time-independent Hamiltonian. 

In a physical scenario when the dressing field amplitude is not constant but slowly changes in time, those matrix elements of $\mathbf{H}_{\rm F}$ which represent light-matter couplings also slowly change in time. For a dressing-field which is turned on much slower than the characteristic timescales of the field-free system (adiabatically), Eq. (\ref{eq:Floquet_temporal_evolution}) suggests that the well-known adiabatic theorem could be used to predict the temporal changes in $\boldsymbol{\Psi} (t)$. 
Therefore, in the limit of the dressing light intensity going to zero ($\mathbf{g}_{\rm XA} \rightarrow 0$) the field-free eigenstates are eigenstates of $\mathbf{H}_{\rm F}$ as well, therefore, an adiabatic turn-on of the dressing field will convert an initial field-free eigenstate into a single light-dressed state. This means that light-dressed states and field-free states can be correlated in a one-to-one fashion. However, it is important to mention that the previous two sentences are not true if $\omega_1$ is in exact resonance with an allowed transition, because this results in the field-free eigenstates not being eigenstates of $\mathbf{H}_{\rm F}$ (but being a linear combination of $\mathbf{H}_{\rm F}$ eigenstates) even for infinitezimal light-matter coupling strengths. Further information on the temporal evolution of light-dressed states in the Floquet formalism can be found for example in Ref. \cite{AdiabaticFloquet2_Gurin_AdvChemPhys_2003}, while Refs. \cite{AdiabaticFloquet_Lefebvre_PRA_2013, AdiabaticFloquet3_Leclerc_PRA_2016,Ammonia_Fabri_JCP_2019,17Zhang} provide examples for the utilization of the adiabatic Floquet dynamics.

\subsection{Computing transitions between light-dressed states} \label{Appendix_C}

\subsubsection{General considerations} \label{Appendix_C_1}

We assume a molecule interacting with two periodic electric fields. The full Hamiltonian reads
\begin{equation}
    \hat{H}(t)=\hat{H}_{\rm mol} + \hat{W}_1 (t) + \hat{W}_2 (t),
\end{equation}
where $\hat{H}_{\rm mol}$ is the field-free molecular Hamiltonian, and $\hat{W}_1(t)$ and $\hat{W}_2(t)$ account for the interaction between the molecule and the two fields, \textit{i.e.}, in the dipole approximation
\begin{equation}
    \begin{split}
    \hat{W}_1 (t) = - \mathbf{E}_1 \hat{\boldsymbol{\mu}} {\rm cos}(\omega _1 t + \phi)\\
    \hat{W}_2 (t) = - \mathbf{E}_2 \hat{\boldsymbol{\mu}} {\rm cos}(\omega _2 t).
    \end{split}
    \label{eq:electric_fields}
\end{equation}
$\hat{W}_1 (t)$ is considered to be generating the light-dressed states, while $\hat{W}_2 (t)$ originates from a weak probe pulse used to record the spectrum of the light-dressed molecule.
The formation of light-dressed states by $\hat{W}_1 (t)$ is accounted for within the Floquet approach, as described in Section \ref{Appendix_A}.

For computing the $\hat{W}_2 (t)$-induced transition amplitudes between the (superposition of) light-dressed states, first-order time-dependent perturbation theory (TDPT1) is used. To derive our working equations, we start with the TDSE containing the interaction with both the dressing and the probe fields,
\begin{equation}
    i \hbar \partial _t \vert \Psi (t) \rangle = (\hat{H}_{\rm mol} + \hat{W}_1 (t) + \hat{W}_2 (t)) \vert \Psi (t) \rangle = ( \hat{H}_{\rm d}(t) + \hat{W}_2 (t)) \vert \Psi (t) \rangle .
    \label{eq:TDSE_Htot}
\end{equation}
Eq. (\ref{eq:TDSE_Htot}) is transformed to the interaction picture using the transformation
\begin{equation}
    \vert \Psi _{\rm I} (t) \rangle = e^{\frac{i}{\hbar}\int _{t_0}^{t} \hat{H}_{\rm d}(t') dt'} \vert \Psi (t) \rangle ,
\end{equation}
which leads to
\begin{equation}
        i \hbar \partial _t \vert \Psi _{\rm I} (t) \rangle = \hat{W}_{2{\rm I}} (t) \vert \Psi _{\rm I} (t) \rangle ,
    \label{eq:TDSE_Htot_I}
\end{equation}
where $ \hat{W}_{2{\rm I}} (t) = e^{\frac{i}{\hbar}\int _{t_0}^{t} \hat{H}_{\rm d}(t') dt'} \hat{W}_2 (t) e^{-\frac{i}{\hbar}\int _{t_0}^{t} \hat{H}_{\rm d}(t') dt'} $. Following the usual TDPT1 procedure of integrating Eq. (\ref{eq:TDSE_Htot_I}) from $t_0$ to $t$ and applying a successive approximation to express $\vert \Psi _{\rm I} (t) \rangle$ gives
\begin{equation}
    \vert \Psi _{\rm I} (t) \rangle = \hat{U}(t,t_0) \vert \Psi _{\rm I} (t_0) \rangle
\end{equation}
with
\begin{equation}
    \hat{U}(t,t_0) = \hat{I} + \frac{1}{i \hbar} \int _{t_0}^{t} \hat{W}_{2{\rm I}} (t') dt' + \frac{1}{(i \hbar)^2}  \int _{t_0}^{t} \hat{W}_{2{\rm I}} (t')  \int _{t_0}^{t'} \hat{W}_{2{\rm I}} (t'') dt'' dt' + \cdots .
    \label{eq:propagator}
\end{equation}
Considering the first two terms of the propagator in Eq. (\ref{eq:propagator}) leads to 
\begin{equation}
    \vert \Psi _{\rm I} (t) \rangle = \vert \Psi _{\rm I} (t_0) \rangle + \frac{1}{i \hbar} \int _{t_0}^{t} \hat{W}_{2{\rm I}} (t') dt' \vert \Psi _{\rm I} (t_0) \rangle .
\end{equation}
The transition amplitude to a final state $\vert \Psi ^{\rm (F)}_{\rm I} \rangle $ at time $t$ is thus
\begin{equation}
    \langle \Psi ^{\rm (F)}_{\rm I} \vert \Psi _{\rm I} (t) \rangle = \langle \Psi ^{\rm (F)}_{\rm I} \vert \Psi _{\rm I} (t_0) \rangle + \frac{1}{i \hbar} \int _{t_0}^{t} \langle \Psi ^{\rm (F)}_{\rm I} \vert \hat{W}_{2{\rm I}} (t') \vert \Psi _{\rm I} (t_0) \rangle dt' .
    \label{eq:transition_amplitude_general}
\end{equation}
By expanding $\vert \Psi ^{\rm (F)}_{\rm I} \rangle$ and $\vert \Psi _{\rm I} (t_0) \rangle$ as a superposition of the $\vert \Phi _k (t_0) \rangle$ Floquet states, \textit{i.e.},
\begin{equation}
    \vert \Psi ^{\rm (F)}_{\rm I} \rangle = \sum_{l} a_l e^{- \frac{i}{\hbar} \varepsilon _l t_0} \vert \Phi _l (t_0) \rangle
\end{equation}
and
\begin{equation}
    \vert \Psi _{\rm I}(t_0) \rangle = \sum_{k} b_k e^{- \frac{i}{\hbar} \varepsilon _k t_0} \vert \Phi _k (t_0) \rangle ,
\end{equation}
Eq. (\ref{eq:transition_amplitude_general}) gives
\begin{equation}
    \begin{split}
        \langle \Psi ^{\rm (F)}_{\rm I} \vert \Psi _{\rm I} (t) \rangle = \\ 
        \sum_{l,k} a^*_l b_k e^{- \frac{i}{\hbar} (\varepsilon _k - \varepsilon _l) t_0} \langle \Phi _{l} (t_0) \vert \Phi _{k} (t_0) \rangle + \frac{1}{i \hbar} \sum_{l,k} a^*_l b_k \int _{t_0}^{t} e^{- \frac{i}{\hbar} (\varepsilon _k - \varepsilon _l) t_0} \langle \Phi _{l} (t_0) \vert \hat{W}_{2{\rm I}} (t') \vert \Phi _{k} (t_0) \rangle dt' = \\ 
        \sum_{l,k} a^*_l b_k e^{- \frac{i}{\hbar} (\varepsilon _k - \varepsilon _l) t_0} \langle \Phi _{l} (t_0) \vert \Phi _{k} (t_0) \rangle + \\ 
        \frac{1}{i \hbar} \sum_{l,k} a^*_l b_k \int _{t_0}^{t} e^{- \frac{i}{\hbar} (\varepsilon _k - \varepsilon _l) t_0} \langle \Phi _{l} (t_0) \vert e^{\frac{i}{\hbar}\int _{t_0}^{t'} \hat{H}_{\rm d}(t'') dt''} \hat{W}_2 (t') e^{-\frac{i}{\hbar}\int _{t_0}^{t'} \hat{H}_{\rm d}(t'') dt''} \vert \Phi _{k} (t_0) \rangle dt' .    
    \end{split}
\end{equation}
Because $e^{- \frac{i}{\hbar} \varepsilon _k t} \vert \Phi _{k} (t) \rangle \Large$ is a solution of the TDSE of Eq. (\ref{eq:Hd_TDSE}), the effect of the $\hat{U}_{\rm d}(t',t_0)=e^{-\frac{i}{\hbar}\int _{t_0}^{t'} \hat{H}_{\rm d}(t'') dt''}$ operator, describing $\hat{H}_{\rm d}(t)$ governed time-evolution from $t_0$ to $t'$, can be evaluated as $\hat{U}_{\rm d}(t',t_0) \Large ( e^{- \frac{i}{\hbar} \varepsilon _k t_0} \vert \Phi _{k} (t_0) \rangle \Large )= e^{-\frac{i}{\hbar} \varepsilon_k t' } \vert \Phi _{k} (t') \rangle$. By exploiting this fact, one arrives at
\begin{equation}
    \begin{split}
        \langle \Psi ^{\rm (F)}_{\rm I} \vert \Psi _{\rm I} (t) \rangle = \\ 
        \sum_{l,k} a^*_l b_k e^{- \frac{i}{\hbar} (\varepsilon _k - \varepsilon _l) t_0} \langle \Phi _{l} (t_0) \vert \Phi _{k} (t_0) \rangle +  
        \frac{1}{i \hbar} \sum_{l,k} a^*_l b_k \int _{t_0}^{t} \langle \Phi _{l} (t') \vert \hat{W}_2 (t') \vert \Phi _{k} (t') \rangle e^{-\frac{i}{\hbar} ( \varepsilon _k - \varepsilon _l) t'} dt' .    
    \end{split}
\end{equation}
Finally, using the explicit form of $ \hat{W}_2 (t) $ given in Eq. (\ref{eq:electric_fields}) and expressing the periodic $\vert \Phi _{k} (t) \rangle$ functions under the integral with their Fourier series (see Eq. (\ref{eq:Floquet_state_Fourier_series})), leads to
\begin{equation}
    \begin{split}
        \langle \Psi ^{\rm (F)}_{\rm I} \vert \Psi _{\rm I} (t) \rangle =  
        \sum_{l,k} a^*_l b_k e^{- \frac{i}{\hbar} (\varepsilon _k - \varepsilon _l) t_0} \langle \Phi _{l} (t_0) \vert \Phi _{k} (t_0) \rangle - \\
        \frac{1}{2 i \hbar} \sum_{l,k} a^*_l b_k \sum_{n,m} \int _{t_0}^{t} \langle \varphi _{ln} \vert \mathbf{E}_2 \hat{\boldsymbol{\mu}} \vert \varphi _{km} \rangle e^{-\frac{i}{\hbar} (\hbar \omega_1 ( n - m ) + \varepsilon _k - \varepsilon _l \pm \hbar \omega_2 ) t'} dt' .    
    \end{split}
    \label{eq:transition_amplitude_Floquet_general}
\end{equation}

\subsubsection{Molecules with no permanent dipole} \label{Appendix_C_2}

For molecules with no permanent dipole and neglectable intrinsic nonadiabatic couplings the light-dressed states determined within a Floquet approach, in which nonresonant coupling terms with the dressing field are neglected, see Eq. (\ref{eq:HF_detailed_2x2}), Eq. (\ref{eq:Floquet_state_Fourier_series}) can be written as 
\begin{equation}
    \vert \Phi _{k} (t) \rangle = \sum_{n} ( \vert \alpha _{kn} \rangle + \vert \beta _{k(n-1)} \rangle e^{-i\omega_1 t} ) e^{in\omega_1 t},
    \label{eq:Floquet_state_Fourier_series_homonucleardiatomic_1}
\end{equation}
where $\vert \alpha _{kn} \rangle $ and $ \vert \beta _{k(n-1)} \rangle$ represent the two manifolds of rovibronic states with Fourier indices $n$ and $n-1$, respectively. Because neglecting the nonresonant coupling terms in the Floquet approach means that the Floquet Hamiltonian becomes block diagonal, see Eq. (\ref{eq:HF_detailed_2x2}), and that $\vert \alpha _{kn} \rangle $ and $ \vert \beta _{k(n-1)} \rangle $ become the same for all $n$, Eq. (\ref{eq:Floquet_state_Fourier_series_homonucleardiatomic_1}) simplifies to
\begin{equation}
    \vert \Phi _{k} (t) \rangle = ( \vert \alpha _{k} \rangle + \vert \beta _{k} \rangle e^{-i\omega_1 t} ) e^{in\omega_1 t},
    \label{eq:Floquet_state_Fourier_series_homonucleardiatomic_2}
\end{equation}
where $\vert \Phi _{k} (t) \rangle$ is a Floquet state obtained from the \textit{n}th two-by-two block of the Floquet Hamiltonian, and quasienergy corresponding to $\vert \Phi _{k} (t) \rangle$ may be written as $\varepsilon_k + \hbar n \omega_1$. 
As explained under Eq. (\ref{eq:Floquet_state_eigenvalue_problem}), if the quasienergy is shifted by $-\hbar n \omega_1$ and the Floquet state is multiplied with $e^{-i n \omega_1 t}$ one arrives to an equivalent physical state. Therefore, Eq. (\ref{eq:Floquet_state_Fourier_series_homonucleardiatomic_2}) can be rewritten as
\begin{equation}
    \vert \Phi _{k} (t) \rangle = \vert \alpha _{k} \rangle + \vert \beta _{k} \rangle e^{-i\omega_1 t},
    \label{eq:Floquet_state_Fourier_series_homonucleardiatomic_3}
\end{equation}
with the corresponding quasienergy of $\varepsilon_k$.
Using Eq. (\ref{eq:Floquet_state_Fourier_series_homonucleardiatomic_3}) instead of Eq. (\ref{eq:Floquet_state_Fourier_series}) leads to a simplified version of Eq. (\ref{eq:transition_amplitude_Floquet_general}), \textit{i.e.},
\begin{equation}
    \begin{split}
        \langle \Psi ^{\rm (F)}_{\rm I} \vert \Psi _{\rm I} (t) \rangle =  
        \sum_{l,k} a^*_l b_k e^{- \frac{i}{\hbar} (\varepsilon _k - \varepsilon _l) t_0} \langle \Phi _{l} (t_0) \vert \Phi _{k} (t_0) \rangle - \\
        \frac{1}{2 i \hbar} \sum_{l,k} a^*_l b_k \int _{t_0}^{t} \langle \beta _{l} \vert \mathbf{E}_2 \hat{\boldsymbol{\mu}} \vert \alpha _{k} \rangle e^{-\frac{i}{\hbar} (\varepsilon _k - \varepsilon  _l - \hbar \omega_1 \pm \hbar \omega_2 ) t'} dt' - \\
        \frac{1}{2 i \hbar} \sum_{l,k} a^*_l b_k \int _{t_0}^{t} \langle \alpha _{l} \vert \mathbf{E}_2 \hat{\boldsymbol{\mu}} \vert \beta _{k} \rangle e^{-\frac{i}{\hbar} (\varepsilon _k - \varepsilon _l + \hbar \omega_1 \pm \hbar \omega_2 ) t'} dt'
        ,    
    \end{split}
    \label{eq:transition_amplitude_Floquet_homonucleardiatomic}
\end{equation}
where we exploited that $\langle \alpha _{l} \vert \mathbf{E}_2 \hat{\boldsymbol{\mu}} \vert \alpha _{k} \rangle = \langle \beta _{l} \vert \mathbf{E}_2 \hat{\boldsymbol{\mu}} \vert \beta _{k} \rangle = 0$ for a molecule with no permanent dipole. Continuing with the standard TDPT1 procedure, the second and third terms in Eq. (\ref{eq:transition_amplitude_Floquet_homonucleardiatomic}) lead to the two terms of Eq. (\ref{eq:Floquet_state_transition_amplitude_2x2_text_body}).

\bibliography{Na2_manuscript_PRA_classical_field,gen}

\end{document}